\documentclass[conference,compsoc]{IEEEtran}

\usepackage{versions}
\usepackage[inline]{enumitem}
\usepackage{xspace}
\usepackage{microtype}
\usepackage{cite}
\usepackage{eurosym}
\usepackage[flushmargin,hang]{footmisc}
\usepackage{url}
\usepackage{booktabs,multirow,makecell}
\usepackage{graphicx}
\usepackage{subcaption}
\usepackage{soul}
\usepackage{xcolor}
\usepackage{amsmath,amssymb}
\usepackage{graphicx}
\usepackage{multirow}
\usepackage{multicol}
\usepackage[colorlinks=true,citecolor=blue]{hyperref}
\usepackage{tabularx}

\usepackage{pifont}
\newcommand{\cmark}{\ding{51}\xspace}%
\newcommand{\xmark}{\ding{55}\xspace}%
\newcommand{\pmark}{$\sim$\xspace}%

\usepackage{amsthm}
\usepackage{thmtools}
\usepackage[most]{tcolorbox}

\usepackage{array}
\newcolumntype{C}[1]{>{\centering\arraybackslash}p{#1}}

\definecolor{blond}{rgb}{0.98, 0.94, 0.75} 

\declaretheoremstyle[numbered=no, headfont=\bfseries, postheadspace=0pt, headpunct={}]{thmsty} 

\declaretheorem[style=thmsty, name={}]{takeaway} 

\tcolorboxenvironment{takeaway}{enhanced jigsaw, colback=blond, drop shadow, boxrule=0.9pt, boxsep=0.1pt, left=4pt, right=4pt, top=4pt, bottom=4pt}

\newcommand{\method}{\textsc{TME}\xspace}
\newcommand{\titlename}{Trusted Model Environment for Private Semantic Computations}

\usepackage[ruled]{algorithm2e} 
\usepackage{cleveref}             

\Crefname{algocf}{Attestation}{Attestations}

\newenvironment{attdesc}[1][htb]
  {%
   \begin{algorithm}[#1]}{\end{algorithm}}

\newcommand{\modelm}{\ensuremath{\mathcal{M}}\xspace}
\newcommand{\modeMon}{\ensuremath{\mathcal{M}_{mon}}\xspace}
\newcommand{\modePara}{\ensuremath{\mathcal{M}_{para}}\xspace}
\newcommand{\hsh}[1]{\ensuremath{\texttt{h}(#1)}}                 
\newcommand{\mroot}[1]{\ensuremath{\operatorname{MR}(#1)}}        
\newcommand{\signatt}{\ensuremath{\operatorname{Sign}_\textrm{att}}}
\newcommand{\attn}[1]{\texttt{`#1'}}                              

\usepackage{tikz}
\usetikzlibrary{positioning, arrows.meta, calc, fit, backgrounds, shapes.geometric}
\pgfdeclarelayer{bg outer}
\pgfdeclarelayer{bg inner}
\pgfsetlayers{bg outer,bg inner,main}

\definecolor{tmefill}{HTML}{E8F1EA}   
\definecolor{srvfill}{HTML}{FBEDEB}   
\definecolor{srvline}{HTML}{8F3A32}
\definecolor{tmeline}{HTML}{2A5C44}
\definecolor{txtfill}{HTML}{E3EEF8}   
\definecolor{imgfill}{HTML}{ECE6F4}   
\definecolor{txtblue}{HTML}{27567E}
\definecolor{imgpurple}{HTML}{5B4390}
\definecolor{piifill}{HTML}{EBE8E2}   
\definecolor{outfill}{HTML}{FFF6DF}   
\definecolor{piired}{HTML}{B03A28}    

\newcommand{\dataverb}[2][orange!90]{%
  \textbf{\textcolor{#1}{\lstinline[breaklines=false, basicstyle=\ttfamily]{#2}}}%
}

\newcommand{\modelverb}[2][blue!70]{%
  \textbf{\textcolor{#1}{\lstinline[breaklines=false, basicstyle=\ttfamily]{#2}}}%
}

\includeversion{arxiv}
\excludeversion{submit}

\begin{document}

\title{\titlename}

\begin{arxiv}
\author{
    \IEEEauthorblockN{Vasisht Duddu, Xi He}
    \IEEEauthorblockA{Vector Institute and University of Waterloo}
    \IEEEauthorblockA{vasisht.duddu@vectorinstitute.ai, xi.he@uwaterloo.ca}
}  
\end{arxiv}

\maketitle

\begin{abstract}
A private semantic computation primitive enables parties to privately compute over structured and unstructured data that requires understanding its semantics, context, and relationships.
Standard cryptographic primitives (e.g., multiparty computation) do not readily support such computation.
Generative models are well suited for such tasks but typically process data in plaintext, while cryptographic private inference remains inefficient and difficult to scale.
Thus, we need a new primitive for private semantic computation.
We introduce \emph{trusted model environments} (\method), the \emph{first such primitive} that executes generative models inside trusted execution environments (TEEs) while controlling output leakage.
\method is designed to be
\begin{enumerate*}[label={(\roman*)}]
\item \emph{effective} (correctly performs the semantic task);
\item \emph{confidential} (protects computation and sensitive inputs);
\item \emph{utility-preserving} (retains utility on other tasks);
\item \emph{verifiable} (provides tamper-resistant evidence of the computations);
\item \emph{efficient} (incurs low overhead compared to baseline model computations); and
\item \emph{scalable} (supports multiple participating parties).
\end{enumerate*}
Effectiveness follows from the generative models, while TEEs provide confidential computation.
For confidentiality of sensitive inputs, we combine adversarial training to resist verbatim leakage with an \emph{information flow control module} to suppress semantic leakage. 
For verifiability, we introduce novel attestations that let parties verify \method operations on their data and queries, along with optimizations (e.g., batching) for efficiency and scalability.
We design and evaluate the proof-of-concept for \method across \emph{three applications}, showing that it meets all the requirements.
\end{abstract}

\section{Introduction}\label{sec:introduction}

The widespread availability of structured (e.g., tabular) and unstructured (e.g., text, images, and graphs) data has spurred \emph{semantic computation}, which requires interpreting data, context, and relationships to extract insights. Given the potential sensitivity of such data, there is a need for \emph{private semantic computation primitives} that support such computations over sensitive data held by multiple parties. For instance, consider the following motivating examples:
\begin{itemize}[leftmargin=*,topsep=0pt,itemsep=0pt]
    \item Alice holds a sensitive document: ``\emph{54-year-old smoker, exertional chest pain radiating to the left arm, relieved by rest; troponin mildly elevated}''. Bob wishes to privately evaluate a function on the document: ``\emph{Are the described symptoms consistent with stable angina?}''. 
    \item  Alice and Bob each hold a confidential set of flagged bank accounts. Alice's set includes \emph{\{shell company routing funds, customer making sub-\$10k deposits\}}; Bob's set includes \emph{\{front business with pass-through transfers, account splitting deposits below threshold\}}. Neither may disclose their individual sets, but they wish to learn the \emph{types} of illicit patterns their lists share (e.g., \emph{{shell-entity}}) despite not being an exact match.
    \item Alice holds a private database of patient notes, and a physician, Bob, asks ``\emph{Which record describes exertional chest pain relieved by rest, with mildly elevated troponin?}''. Alice needs to retrieve the relevant record without knowing Bob's query or leaking the access pattern.
\end{itemize}
The outputs in all these cases are determined by interpreting the data, context, and relationships.

Standard cryptographic primitives (e.g., private function computation~\cite{BonehSW11,Gentry09,GoldreichMW87,Yao86}, private information retrieval~\cite{ChorKGS98}, oblivious RAM~\cite{Ostrovsky90,StefanovDSFRYD13}, and private set intersection~\cite{FreedmanNP04,VatsalanCV13,HallF10}) are designed for structured data types, and do not readily support semantic computations.
On the other hand, generative models, such as large language models (LLMs)~\cite{VaswaniSPUJGKP17,BrownMRSKDNSSAA20} and vision-language models (VLMs)~\cite{RadfordKHRGASAM21,AlayracDLMBHLMM22}, are effective for semantic computation over structured and unstructured data. 
However, they operate on the data in plaintext and do not provide any confidentiality guarantees. 
Moreover, private inference of generative models using cryptographic primitives is expensive in practice~\cite{iron,ciphergpt,bumblebee,bolt,puma,sigma,thor,shaft,nexus}.
Thus, we need a new private semantic computation primitive.


Trusted execution environments (TEEs) can efficiently run generative models while offering computation confidentiality~\cite{chrapek2024fortify,chantasantitam2026pal,petridish,talaria,cmif} and verifiability~\cite{chantasantitam2026pal,laminator}. Thus, TEEs are a promising alternative, but simply running models inside TEEs is insufficient as the models can still leak sensitive inputs through their outputs (\S\ref{sec:evaluation}).
We propose \emph{trusted model environments} (\method), which execute generative models inside TEEs while meeting the following requirements:
\begin{enumerate*}[label={(\roman*)}]
\item \emph{effective} (correctly performs the semantic task);
\item \emph{confidential} (protects computation and sensitive inputs)\footnote{\textbf{Terminology:} We treat computation and sensitive-input confidentiality separately under a unified ``confidentiality'' requirement. While other definitions are possible, we define sensitive-input confidentiality as preventing sensitive inputs from appearing in outputs, following prior work on personally identifiable information leakage in LLMs~\cite{ponomarenko}.};
\item \emph{utility-preserving} (retains utility on other tasks);
\item \emph{verifiable} (provides tamper-resistant evidence of the computations);
\item \emph{efficient} (incurs low overhead compared to baseline model computations); and
\item \emph{scalable} (supports multiple parties).
\end{enumerate*}
Effectiveness is inherited from the generative models while TEEs provide confidential computation.
However, to ensure confidentiality of sensitive inputs, we adapt latent adversarial training (LAT)~\cite{sheshadri2024latent} to suppress verbatim leakage of sensitive inputs while preserving the model's effectiveness and utility.
To mitigate semantic leakage caused by paraphrasing sensitive inputs in the output, an \emph{information flow control (IFC) module} monitors outputs and paraphrases those flagged as containing sensitive information.
Additionally, due to the lack of transparency into \method, parties may wish to verify the computations on their data. 
For this, we build on remote attestation of TEEs and propose novel attestations for various \method operations.
We additionally introduce batching and amortization to improve efficiency and support multiple parties.
We claim the following contributions:
\begin{enumerate}[leftmargin=*,topsep=0pt,itemsep=0pt]
\item identify the requirements for an ideal private semantic computation primitive, and highlight the limitations of existing approaches; (\S\ref{sec:problem})
\item present trusted model environments (\method), the first design and implementation of such a primitive supporting structured and unstructured data across modalities; (\S\ref{sec:approach})
\item demonstrate \method on the three applications\footnote{Code will be open-sourced upon publication.}, showing that \method is \emph{effective} (close to the base model), \emph{confidential} (negligible verbatim and semantic leakage), \emph{utility-preserving} (drop $<6$pp), and \emph{efficiently} supports \emph{verifiability} across multiple parties (\S\ref{sec:setup} and \S\ref{sec:evaluation}).
\end{enumerate}

\section{Background}\label{sec:background}

\noindent\textbf{\ul{Cryptographic Primitives:}} We describe three cryptographic primitives, which will inform the applications to demonstrate our primitive (\S\ref{sec:problem}).
\begin{itemize}[leftmargin=*,topsep=0pt,itemsep=0pt]
    \item \textbf{Private Function Computation:} In \emph{secure multi-party computation}, the parties jointly compute a function $f$ over their private inputs without revealing anything beyond the output~\cite{Yao86,GoldreichMW87}. In \emph{homomorphic encryption}, party-1 encrypts the private input $x$ while party-2 evaluates $f$ on the ciphertext, and only party-1 can decrypt the result~\cite{Gentry09}. In \emph{functional encryption}, party-2 receives a function-specific secret key that enables it to learn only $f(x)$ from party-1's ciphertext, revealing nothing else about $x$~\cite{BonehSW11}. These primitives differ in who performs the computation and who learns the output. 
    \item \textbf{Private Set Intersection:} Two parties each hold a private set of elements and wish to jointly compute the intersection such that (at least) one party learns the intersection, while neither party learns anything about the set elements~\cite{FreedmanNP04}. Private record linkage extends this to structured records by identifying records that refer to the same real-world entity using approximate (fuzzy) matching. Unlike exact set intersection, this tolerates noisy or inconsistently formatted data while revealing only the matched records~\cite{VatsalanCV13,HallF10}.
    \item \textbf{Private Database Retrieval:} An untrusted server hosts a private database of documents, and various parties can query the server to retrieve the relevant entries. While encryption hides the records' contents, the \emph{access pattern} (which entries each query accesses) can still leak sensitive information about both the query and the data. Oblivious RAM allows querying the private database without revealing the access pattern of the records being retrieved~\cite{Ostrovsky90,StefanovDSFRYD13}. This differs from private information retrieval~\cite{ChorKGS98}, where a client retrieves an entry from a server-held (often public, read-only) database without the server learning \emph{which} one.
\end{itemize}

\noindent\textbf{\ul{Trusted Execution Environments (TEEs):}} TEEs provide an environment in which security-critical code (trusted applications or TAs) can run without outside interference (e.g., untrusted operating system and software).
TEEs have three main properties: \emph{isolation} and \emph{protected memory} for confidentiality and integrity, and \emph{remote attestation}~\cite{gunn2022hardware}.
\begin{itemize}[leftmargin=*,topsep=0pt,itemsep=0pt]
    \item \emph{Isolation} is the inability of outside software to read or write a TA. TEEs provide strong isolation between trusted and untrusted components. Thus, code and data in the TEE are protected from external software, and access is limited to well-defined interfaces.
    \item \emph{Protected memory} is the encrypted and integrity-protected memory available to TAs, to ensure that sensitive data cannot be read or modified by untrusted components.
    \item \emph{Remote attestation} allows a prover to convince a verifier that a specific program is running inside a genuine TEE. The TEE computes a measurement and produces an attestation $\sigma = \mathsf{Sign}_{sk_{\text{TEE}}}(\mathsf{Hash}(Code), \text{nonce})$, where $sk_{\text{TEE}}$ is a hardware-protected secret attestation key for signing, and $\text{nonce}$ is a fresh random challenge chosen by the verifier for each attestation request, whose inclusion proves the attestation was generated for the current request, and prevents replay of an old one. The verifier checks whether $\mathsf{Verify}_{pk_{\text{TEE}}}(\sigma, \mathsf{Hash}(Code), \text{nonce}) = 1$, where $pk_{\text{TEE}}$ is the public key. This ensures authenticity (genuine TEE) and integrity (attests correct code execution).
\end{itemize}
Examples of TEEs include Intel SGX, Intel TDX (supports GPUs), and AMD SEV.
Recent advances in TEEs, including hardware acceleration of matrix multiplication for generative models with GPU support, enable efficient execution while retaining TEE guarantees. These capabilities are readily available through major cloud providers such as Google Cloud, Microsoft Azure, and Amazon Web Services. Together, they make it practical to run large generative models inside TEEs. We discuss TEE side-channel attacks in \S\ref{sec:discussion}.


\noindent\textbf{\ul{Generative Models:}} A generative model is a function $f$ that maps an input $x$ to an output $y \leftarrow f(x)$. Depending on the model type, both $x$ and $y$ may be text, images, or a combination of the two. We focus on LLMs, which have demonstrated remarkable utility in natural language understanding, reasoning, and text generation by scaling transformer-based architectures and training on massive text corpora. LLMs support efficient modeling of long-range dependencies, while increasing model and data size improves utility on downstream tasks~\cite{VaswaniSPUJGKP17,BrownMRSKDNSSAA20}.
We use LLMs for our evaluation (\S\ref{sec:evaluation}), and discuss other modalities (\S\ref{sec:discussion}).
\section{Problem Statement}\label{sec:problem}

Our goal is to design a private computation primitive which enables semantic computations on structured and unstructured data. 

\subsection{Threat Model}\label{sec:threatModel}

We consider a general setting for \emph{semantic computation}, in which multiple parties wish to evaluate semantic functions or queries over privately held data containing sensitive information (e.g., personally identifiable information such as race and occupation) as defined by the data owners, 
whether supplied directly as one party's inputs or stored in a database.
The goal is to furnish the correct responses to the queries from various parties without revealing sensitive information (e.g., parties' sensitive inputs containing personally identifiable information such as race or address), even under adversarial queries.
The computations are executed on an untrusted server.
As the parties do not trust the server or one another, the sensitive inputs, intermediate computation, and outputs must be protected.

We assume that the server runs a TEE that is trusted by all parties, while all software outside the TEE (e.g., operating system, hypervisor, and other applications) is untrusted. Through remote attestation, parties can verify that the expected code runs unmodified on genuine TEE hardware, without trusting the server. The parties trust the hardware manufacturer (e.g., Intel or Nvidia), and any attestation signed using their attestation key is trusted. Side-channel attacks against TEEs are outside our scope (discussed in \S\ref{sec:discussion}).

We assume that one or more parties can be (a) an \emph{honest-but-curious adversary}, who follows the protocol correctly but attempts to infer additional information about other parties' sensitive inputs from the messages observed during execution; or (b) a \emph{malicious adversary}, who may deviate from the protocol to compromise privacy or correctness by sending malformed messages or using adversarial queries.
We assume that the adversary knows the defense and uses the feedback from the target model to optimize subsequent attacks and queries (or ``adaptive adversaries''; see Appendix~\ref{app:adaptive}). 
This is a worst-case robustness evaluation as the defenses optimized against such attacks should withstand weaker attacks seen in practice. 

\subsection{Illustrative Applications}\label{sec:applications}


We discuss three illustrative applications: (i) computing a semantic function specified by some parties over another party’s private input (PSFC), (ii) computing the semantic intersection of several parties’ private sets (PSSP), and (iii) semantically retrieving records from a private database of documents based on queries (PSDR).
 \begin{itemize}[leftmargin=*,topsep=0pt,itemsep=0pt]
    \item \textbf{\emph{\ul{Private Semantic Function Computation (PSFC):}}} Alice holds a private input $x$ (e.g., a document), part of which includes sensitive inputs, and $N$ mutually distrusting parties each hold a private function $f_i$ they wish to evaluate over $x$. Party $i$ ($\neq$Alice) should learn $f_i(x)$ and not the sensitive inputs. Private function computation primitives (\S\ref{sec:background}) express $f$ as an arithmetic or Boolean circuit and are thus confined to predicates such as comparison, threshold, or exact match. Here, we instead instantiate each $f_i$ as a \emph{free-text reading-comprehension question} (semantic, open-ended, and not reducible to such a circuit) while evaluating the effectiveness by comparing the responses with benchmark ground truth answers.
    
   \textbf{Example:} Alice's document reads: ``\emph{The Old English dialects took shape as the early kingdoms settled. Northumbria lay south of the Tyne, and Mercia held the midlands, until both were overrun by Vikings in the ninth century. Only Early West Saxon survived as a literary standard, and under it much religious writing was translated from Latin into the vernacular.}''
    \begin{itemize}[leftmargin=*,topsep=0pt,itemsep=0pt]
      \item \textbf{Sensitive Inputs:} \emph{Tyne}, \emph{Mercia}, and \emph{Early West Saxon}.
      \item \textbf{Query}: ``\emph{From what language was literature translated into West Saxon?}'' $\rightarrow$ ``\emph{Latin}''. 
      \item \textbf{Query}: ``\emph{In which century did the Viking conquest occur?}'' $\rightarrow$ ``\emph{Ninth century}''.
    \end{itemize}

\item \textbf{\emph{\ul{Private Semantic Set Processing (PSSP):}}} Consider $N$ mutually distrusting parties, each holding a private set $S_i$ of records (e.g., documents or images). 
Every party should learn only the semantic categories across the sets and nothing more (without revealing the sets held by the parties). 
Standard private set intersection primitives (\S\ref{sec:background}) compute the intersection of sets of \emph{exact} elements or close approximations (fuzzy-intersection primitives), and cannot match records by meaning. Here, the output is inferred from the semantics underlying various set elements, not reducible to exact-element equality.

  \textbf{Example:} Sets include various prescribed drugs, and a query: ``\emph{Which drug classes do all parties prescribe that share a bleeding-risk contraindication?}''
    \begin{itemize}[leftmargin=*,topsep=0pt,itemsep=0pt]
      \item \textbf{Sets:} Alice $=\{$warfarin, aspirin$\}$; Bob $=\{$clopidogrel, naproxen$\}$; Charlie $=\{$heparin, diclofenac$\}$.
      \item \textbf{Response:} ``\emph{Anticoagulants and NSAIDs}''.
    \end{itemize}

\item \textbf{\emph{\ul{Private Semantic Database Retrieval (PSDR):}}} Alice holds a private database of $N_{db}$ records (e.g., text passages or images), and $N$ parties each hold a private query $q_j$ whose answer lies in some subset of the records. Party $j$ should learn only the answer to $q_j$, and the untrusted server observing the computation should learn nothing more about the private database (i.e., no \emph{access pattern} leakage) or $q_j$. 
Private database retrieval primitives (\S\ref{sec:background}) hide the access pattern and the query index, but retrieve records by position or exact keyword/similarity match, and return the records themselves rather than an answer. 
Here, we instead instantiate each $q_j$ as a \emph{free-text semantic query answered over the records} (open-ended queries whose responses are distilled from the database records, not reducible to an index or similarity lookup).
For instance, consider the PSFC example above, where Alice's document is now one of many records in the database. Given a query, the goal is to identify the closest document and distill the answer from it, without revealing the access pattern or the sensitive inputs in the documents. 

\end{itemize}
We want the outputs not to disclose sensitive inputs (i.e., sets) while correctly responding to task-specific queries.

\subsection{Limitations of Related Work}\label{sec:limitations}
We discuss some possible primitives to realize private semantic computation, and their limitations.

\noindent\textbf{Standard cryptographic primitives} operate on structured data covering specific data types (e.g., fixed-length numeric vectors and matrices, integers and finite-field elements for arithmetic circuits, Booleans or keyword sets). Thus, these primitives are not directly applicable to the wide range of semantic computation on unstructured data. 
Extensions of these primitives that support private computation based on metadata or record similarity have been proposed~\cite{zhou2025pacmann,sanns,wally,chor-keywordpir,coeus,ipfe,keytoindex-pir,cw-pir}. However, their high overheads limit practical deployment, and they cannot directly support the semantic computations required in our applications (\S\ref{sec:applications}).

\noindent\textbf{Generative models} (e.g., LLMs) are highly effective at semantic computation, powering chatbots~\cite{achiam2023gpt,team2023gemini,grattafiori2024llama}, retrieval-augmented generation (RAG)~\cite{lewis2020retrieval}, and autonomous agents~\cite{openai_codex,anthropic_claude}. However, state-of-the-art open-weight models provide no confidentiality of the computation. Additionally, by default, the models can be forced to reveal sensitive inputs, requiring defenses tailored to our settings. 

\noindent\textbf{Private inference} of generative models (e.g., LLMs) can enable semantic computation with confidentiality, using primitives such as secure multi-party computation, zero-knowledge proofs, and homomorphic encryption~\cite{iron,ciphergpt,bumblebee,bolt,puma,sigma,thor,shaft,nexus,zkllm}. However, the reported overheads for cryptographic private inference are in the hundreds of seconds per token, which is expensive.
Several frameworks use cryptographic primitives for privacy-preserving RAG (a problem similar to PSDR)~\cite{pir-rag,prag,sag}, but incur a high overhead.


\noindent\textbf{Other related work} includes $\pi$Creds~\cite{breckenridge2026pi}, which is a privacy-preserving decentralized verifiable credential system that runs LLM inference inside a TEE over authenticated unstructured data to enable semantic claims. It targets a \emph{different use case}, and does not support the applications we want to explore (PSFC, PSSP, and PSDR; \S\ref{sec:applications}).
Finally, the closest related work is by Shumailov et al.~\cite{shumailov2025trusted}, who propose \emph{trusted capable model environments} by running LLMs inside TEEs and indicate the need for attestation and model robustness.
While conceptually similar, their work describes the idea and its potential, but does not specify concrete details for realizing them. 
We are the first to design and implement the necessary components to realize such a primitive.

\begin{takeaway}
\noindent\textbf{Summary:} Existing approaches have several limitations, and cannot directly be used to realize the primitive.
\end{takeaway}
\section{Trusted Model Environment}\label{sec:approach}

\begin{figure*}[!htbp]
\centering
\begin{subfigure}[t]{0.32\textwidth}
  \centering
  \resizebox{!}{1.29\linewidth}{\begin{tikzpicture}[
  comp/.style={draw, line width=0.8pt, fill=black!10,
               rounded corners=2.5pt,
               align=center, font=\bfseries\footnotesize,
               inner ysep=4pt, minimum height=0.92cm},
  bar/.style={comp, inner ysep=3.5pt, minimum height=0pt},
  panel/.style={fill=txtfill, draw=black!45, line width=0.5pt,
                rounded corners=2.5pt,
                align=left, font=\footnotesize, text width=7.8cm,
                inner xsep=5pt, inner ysep=4pt},
  qbox/.style={fill=txtfill, draw=black!45, line width=0.5pt,
               rounded corners=2.5pt, align=center, font=\footnotesize, text width=7.3cm,
               inner xsep=5pt, inner ysep=3pt,
               execute at begin node={\hyphenpenalty=10000\exhyphenpenalty=10000\relax}},
  card/.style={draw=black!70, fill=white, line width=0.6pt,
               rounded corners=2.5pt, align=center, font=\bfseries\footnotesize,
               inner xsep=6pt, inner ysep=4pt},
  snum/.style={circle, fill=black, text=white,
               inner sep=1.4pt, font=\bfseries\footnotesize,
               minimum size=12pt},
  flow/.style={-{Stealth[length=8pt]}, line width=2.2pt},
  flow2/.style={{Stealth[length=8pt]}-{Stealth[length=8pt]}, line width=2.2pt},
  hw/.style={{Stealth[length=5pt]}-{Stealth[length=5pt]}, line width=1.3pt},
  measbox/.style={draw, line width=0.8pt, fill=blond,
                  rounded corners=2.5pt, align=center,
                  font=\bfseries\scriptsize, inner ysep=2.5pt},
]

\node[comp, text width=2.05cm, minimum height=1.30cm] (llm) {Model (LLM)};
\node[comp, text width=3.2cm, right=0.55cm of llm]
  (mon) {Monitor + Paraphraser\\ {\mdseries\footnotesize (Information Flow Control Module)}};

\coordinate (barx) at ($(llm.west)!0.5!(mon.east)$);

\node[measbox, text width=2.05cm, anchor=north] (measl) at (llm.south)
  {Measurer Script};
\node[measbox, text width=3.2cm, anchor=north] (measr) at (mon.south)
  {Measurer Script};

\node[font=\bfseries\footnotesize, text=tmeline, align=center, anchor=south] (tmetitle)
  at ($(barx |- mon.north) + (0,0.22)$)
  {Trusted Model Environment (TME)\\
   {\mdseries\footnotesize Intel Trust Domain running Guest OS}};

\begin{pgfonlayer}{bg inner}
\node[draw=tmeline, line width=0.9pt, fill=tmefill,
      rounded corners=3.5pt, inner sep=6pt, inner ysep=7pt,
      fit={(llm) (mon) (measl) (measr) (tmetitle)}] (tme) {};
\end{pgfonlayer}

\draw[flow] (llm.east) -- (mon.west) node[snum, midway] {3};

\node[comp, rotate=90, minimum width=1.6cm, minimum height=0pt, inner ysep=2.5pt,
      font=\bfseries\footnotesize]
  (tdx) at ($(tme.east)+(0.70,1.0)$) {TDX Module};
\node[comp, rotate=90, minimum width=1.6cm, minimum height=0pt, inner ysep=2.5pt,
      font=\bfseries\footnotesize]
  (h100) at ($(tme.east)+(0.70,-1.0)$) {H100 GPU};
\draw[hw] (tdx.north) -- (tdx.north -| tme.east);
\draw[hw] (h100.north) -- (h100.north -| tme.east);

\begin{pgfonlayer}{bg outer}
\node[draw=srvline, line width=0.9pt, fill=srvfill,
      rounded corners=3.5pt, inner sep=6pt, inner ysep=6pt,
      fit={(tme) (tdx) (h100) ($(tme.south)+(0,-0.32)$)}] (srv) {};
\end{pgfonlayer}
\node[font=\bfseries\footnotesize, text=srvline, anchor=south east,
      inner sep=2.5pt] at (srv.south east) {Untrusted Server};

\path let \p1=(srv.west), \p2=(srv.east) in
  node[panel, text width={\x2-\x1-19.6pt}, anchor=south] (extext)
  at ($(srv.north)+(0,1.00)$)
  {{\bfseries Private Input (Party 1)}\\[1pt]
   ``Each of these dialects was [\ldots] Northumbria south of the
   {\color{piired}\bfseries Tyne}, and {\color{piired}\bfseries Mercia},
   were overrun by Vikings [\ldots] {\color{piired}\bfseries Early West
   Saxon} [\ldots] translated from Latin.''\\[2pt]
   \colorbox{piifill}{\parbox{\dimexpr\linewidth-2\fboxsep\relax}{\raggedright\scriptsize\ttfamily Sensitive Inputs =
     [{\color{piired}`Tyne'}, {\color{piired}`Mercia'},
      {\color{piired}`Early West Saxon'}]}}};
\draw[flow] (extext.south) -- (extext.south |- tme.north);
\node[snum] at ($(extext.south)!0.5!(extext.south |- srv.north)+(-0.38,0)$) {1};
\node[font=\bfseries\footnotesize, anchor=west]
  at ($(extext.south)!0.5!(extext.south |- srv.north)+(0.30,0)$) {Private Input};

\node[qbox, text width=3.6cm, anchor=north west] (qtext) at ($(srv.west |- srv.south)+(0.16,-1.98)$)
  {From what language was literature
   translated into West Saxon?};
\node[font=\bfseries\footnotesize, anchor=south west] (u1lab)
  at ($(qtext.north west)+(0,0.04)$) {Party 2};
\begin{pgfonlayer}{bg inner}
\node[card, inner sep=4pt, fit={(qtext) (u1lab)}] (u1) {};
\end{pgfonlayer}

\node[card, dashed, inner sep=0pt,
      fit={($(u1.north east)+(0.30,0)$) (u1.south east -| srv.east)}] (uN) {};
\node[font=\bfseries\footnotesize, anchor=north] at ($(uN.north)+(0,-0.10)$)
  {Parties $3,\dots,N$};
\node[font=\Large\bfseries] at ($(uN.center)+(0,-0.12)$) {$\cdots$};
\node[font=\footnotesize, align=center, anchor=south]
  at ($(uN.south)+(0,0.08)$) {(may be\\[-2pt] adversarial)};
\draw[flow, black!65, {Stealth[length=8pt]}-{Stealth[length=8pt]}]
  (uN.north) -- (uN.north |- srv.south);

\coordinate (qx) at ($(u1.north west)!0.32!(u1.north east)$);
\coordinate (rx) at ($(u1.north west)!0.75!(u1.north east)$);
\draw[flow, txtblue]   (qx) -- (qx |- tme.south);
\node[snum] at ($(qx)!0.5!(qx |- srv.south) + (0.68,0)$) {2};
\draw[flow, txtblue]   (rx |- tme.south) -- (rx);
\node[snum] at ($(rx)!0.5!(rx |- srv.south) + (0.68,0)$) {4};

\end{tikzpicture}}
  \caption{Private Semantic Function Computation}
  \label{fig:psfc}
\end{subfigure}
\hfill
\begin{subfigure}[t]{0.32\textwidth}
  \centering
  \resizebox{!}{1.29\linewidth}{\begin{tikzpicture}[
  comp/.style={draw, line width=0.8pt, fill=black!10,
               rounded corners=2.5pt,
               align=center, font=\bfseries\footnotesize,
               inner ysep=4pt, minimum height=0.92cm},
  bar/.style={comp, inner ysep=3.5pt, minimum height=0pt},
  panel/.style={fill=txtfill, draw=black!45, line width=0.5pt,
                rounded corners=2.5pt,
                align=left, font=\footnotesize, text width=9.2cm,
                inner xsep=5pt, inner ysep=4pt,
                execute at begin node={\hyphenpenalty=10000\exhyphenpenalty=10000\relax}},
  chip/.style={fill=txtfill, draw=black!45, line width=0.5pt,
               rounded corners=2.5pt, align=center, font=\footnotesize,
               text width=2.05cm, inner xsep=2.5pt, inner ysep=2pt,
               execute at begin node={\hyphenpenalty=10000\exhyphenpenalty=10000\relax}},
  card/.style={draw=black!70, fill=white, line width=0.6pt,
               rounded corners=2.5pt, align=center, font=\bfseries\footnotesize,
               inner xsep=6pt, inner ysep=4pt},
  snum/.style={circle, fill=black, text=white,
               inner sep=1.4pt, font=\bfseries\footnotesize,
               minimum size=12pt},
  flow/.style={-{Stealth[length=8pt]}, line width=2.2pt},
  flow2/.style={{Stealth[length=8pt]}-{Stealth[length=8pt]}, line width=2.2pt},
  hw/.style={{Stealth[length=5pt]}-{Stealth[length=5pt]}, line width=1.3pt},
  measbox/.style={draw, line width=0.8pt, fill=blond,
                  rounded corners=2.5pt, align=center,
                  font=\bfseries\scriptsize, inner ysep=2.5pt},
]

\node[comp, text width=2.05cm, minimum height=1.30cm] (llm) {Model (LLM)};
\node[comp, text width=3.2cm, right=0.55cm of llm]
  (mon) {Monitor + Paraphraser\\ {\mdseries\footnotesize (Information Flow Control Module)}};

\coordinate (barx) at ($(llm.west)!0.5!(mon.east)$);

\node[font=\bfseries\footnotesize, text=tmeline, align=center, anchor=south] (tmetitle)
  at ($(barx |- mon.north) + (0,0.22)$)
  {Trusted Model Environment (TME)\\
   {\mdseries\footnotesize Intel Trust Domain running Guest OS}};

\node[measbox, text width=2.05cm, anchor=north] (measl) at (llm.south)
  {Measurer Script};
\node[measbox, text width=3.2cm, anchor=north] (measr) at (mon.south)
  {Measurer Script};

\begin{pgfonlayer}{bg inner}
\node[draw=tmeline, line width=0.9pt, fill=tmefill,
      rounded corners=3.5pt, inner sep=6pt, inner ysep=7pt,
      fit={(llm) (mon) (measl) (measr) (tmetitle)}] (tme) {};
\end{pgfonlayer}

\draw[flow] (llm.east) -- (mon.west) node[snum, midway] {3};

\node[comp, rotate=90, minimum width=1.6cm, minimum height=0pt, inner ysep=2.5pt,
      font=\bfseries\footnotesize]
  (tdx) at ($(tme.east)+(0.70,1.0)$) {TDX Module};
\node[comp, rotate=90, minimum width=1.6cm, minimum height=0pt, inner ysep=2.5pt,
      font=\bfseries\footnotesize]
  (h100) at ($(tme.east)+(0.70,-1.0)$) {H100 GPU};
\draw[hw] (tdx.north) -- (tdx.north -| tme.east);
\draw[hw] (h100.north) -- (h100.north -| tme.east);

\begin{pgfonlayer}{bg outer}
\node[draw=srvline, line width=0.9pt, fill=srvfill,
      rounded corners=3.5pt, inner sep=6pt, inner ysep=6pt,
      fit={(tme) (tdx) (h100) ($(tme.south)+(0,-0.32)$)}] (srv) {};
\end{pgfonlayer}
\node[font=\bfseries\footnotesize, text=srvline, anchor=south east,
      inner sep=2.5pt] at (srv.south east) {Untrusted Server};

\path let \p1=(srv.west), \p2=(srv.east) in
  node[panel, text width={\x2-\x1-11pt}, inner ysep=8pt, anchor=south] (systext)
  at ($(srv.north)+(0,0.70)$)
  {{\bfseries System Prompt}\; ``Which drug classes do all three
   prescribe that share a bleeding-risk contraindication?''};
\draw[flow] (systext.south) -- (systext.south |- tme.north);
\node[snum] at ($(systext.south)!0.5!(systext.south |- srv.north)+(-0.38,0)$) {1};

\node[chip, anchor=north west] (at) at ($(srv.west |- srv.south)+(0.16,-1.75)$)
  {\{warfarin \ldots\}};
\node[font=\bfseries\footnotesize, anchor=south west] (alab)
  at ($(at.north west)+(0,0.04)$) {Party A};
\begin{pgfonlayer}{bg inner}
\node[card, inner sep=4pt, fit={(at) (alab)}] (pa) {};
\end{pgfonlayer}

\node[chip, anchor=north] (bt) at ($(srv.south)+(0,-1.75)$)
  {\{clopidogrel \ldots\}};
\node[font=\bfseries\footnotesize, anchor=south west] (blab)
  at ($(bt.north west)+(0,0.04)$) {Party B};
\begin{pgfonlayer}{bg inner}
\node[card, inner sep=4pt, fit={(bt) (blab)}] (pb) {};
\end{pgfonlayer}

\node[chip, anchor=north east] (ct) at ($(srv.east |- srv.south)+(-0.16,-1.75)$)
  {\{heparin \ldots\}};
\node[font=\bfseries\footnotesize, anchor=south west] (clab)
  at ($(ct.north west)+(0,0.04)$) {Party C};
\begin{pgfonlayer}{bg inner}
\node[card, inner sep=4pt, fit={(ct) (clab)}] (pc) {};
\end{pgfonlayer}

\draw[flow, txtblue]   ($(pa.north)+(-0.22,0)$) -- ($(pa.north |- srv.south)+(-0.22,0)$);
\node[snum] at ($(pa.north)!0.5!(pa.north |- srv.south) + (-0.68,0)$) {2};
\draw[flow, txtblue]   ($(pb.north)+(-0.22,0)$) -- ($(pb.north |- srv.south)+(-0.22,0)$);
\draw[flow, txtblue]   ($(pc.north)+(-0.22,0)$) -- ($(pc.north |- srv.south)+(-0.22,0)$);

\path let \p1=(srv.west), \p2=(srv.east) in
  node[chip, font=\footnotesize, text width={\x2-\x1-15pt}, inner ysep=5pt, anchor=north] (outt)
  at ($(pb.north |- pa.south)+(0,-1.38)$)
  {intersection = ``anticoagulants + NSAIDs''};
\node[font=\bfseries\footnotesize, anchor=south west] (outlab)
  at ($(outt.north west)+(0,0.04)$) {Output (broadcast to all parties)};
\begin{pgfonlayer}{bg inner}
\node[card, fill=outfill, inner sep=4pt,
      fit={(outt) (outlab) ($(outt.center -| pa.west)+(0.15,0)$) ($(outt.center -| pc.east)+(-0.15,0)$)}] (out) {};
\end{pgfonlayer}
\node[snum] at ($(out.north east)+(-0.36,-0.30)$) {4};
\end{tikzpicture}}
  \caption{Private Semantic Set Processing}
  \label{fig:pssp}
\end{subfigure}
\hfill
\begin{subfigure}[t]{0.32\textwidth}
  \centering
  \resizebox{!}{1.29\linewidth}{\begin{tikzpicture}[
  comp/.style={draw, line width=0.8pt, fill=black!10,
               rounded corners=2.5pt,
               align=center, font=\bfseries\footnotesize,
               inner ysep=4pt, minimum height=0.92cm},
  bar/.style={comp, inner ysep=3.5pt, minimum height=0pt},
  panel/.style={fill=txtfill, draw=black!45, line width=0.5pt,
                rounded corners=2.5pt,
                align=left, font=\footnotesize, text width=9.2cm,
                inner xsep=5pt, inner ysep=4pt},
  qbox/.style={fill=txtfill, draw=black!45, line width=0.5pt,
               rounded corners=2.5pt, align=center, font=\footnotesize, text width=7.3cm,
               inner xsep=5pt, inner ysep=3pt,
               execute at begin node={\hyphenpenalty=10000\exhyphenpenalty=10000\relax}},
  card/.style={draw=black!70, fill=white, line width=0.6pt,
               rounded corners=2.5pt, align=center, font=\bfseries\footnotesize,
               inner xsep=6pt, inner ysep=4pt},
  snum/.style={circle, fill=black, text=white,
               inner sep=1.4pt, font=\bfseries\footnotesize,
               minimum size=12pt},
  flow/.style={-{Stealth[length=8pt]}, line width=2.2pt},
  flow2/.style={{Stealth[length=8pt]}-{Stealth[length=8pt]}, line width=2.2pt},
  hw/.style={{Stealth[length=5pt]}-{Stealth[length=5pt]}, line width=1.3pt},
  measbox/.style={draw, line width=0.8pt, fill=blond,
                  rounded corners=2.5pt, align=center,
                  font=\bfseries\scriptsize, inner ysep=2.5pt},
]

\node[comp, text width=2.05cm, minimum height=1.30cm] (llm) {Model (LLM)};
\node[comp, text width=3.2cm, right=0.55cm of llm]
  (mon) {Monitor + Paraphraser\\ {\mdseries\footnotesize (Information Flow Control Module)}};

\coordinate (barx) at ($(llm.west)!0.5!(mon.east)$);

\node[cylinder, shape border rotate=90, aspect=0.16, draw, line width=0.8pt,
      fill=black!10, align=center, font=\bfseries\footnotesize,
      inner xsep=3pt, minimum width=2.1cm, anchor=south]
  (db) at ($(llm.north west)+(1.45,0.55)$)
  {Private Database\\ {\mdseries\footnotesize (docs w/ sensitive inputs)}};
\node[comp, text width=1.95cm, minimum height=0.55cm, inner ysep=2.5pt, anchor=east]
  (car) at ($(mon.east |- db.center)+(0,0.18)$) {Carousel};
\node[measbox, text width=1.95cm, anchor=north] (measc) at (car.south)
  {Measurer Script};
\draw[flow] (db.east) -- (db.east -| car.west)
  node[snum, midway, yshift=0.34cm] {3};
\draw[flow] ($(measc.south west)+(0.9,0)$) .. controls +(0,-0.30) and +(0.5,0.40) ..
  ($(llm.north east)+(-0.45,0)$) node[snum, pos=0.5] {4};

\node[font=\bfseries\footnotesize, text=tmeline, align=center, anchor=south] (tmetitle)
  at ($(barx |- db.north) + (0.35,0.16)$)
  {Trusted Model Environment (TME)\\
   {\mdseries\footnotesize Intel Trust Domain running Guest OS}};

\node[measbox, text width=2.05cm, anchor=north] (measl) at (llm.south)
  {Measurer Script};
\node[measbox, text width=3.2cm, anchor=north] (measr) at (mon.south)
  {Measurer Script};

\begin{pgfonlayer}{bg inner}
\node[draw=tmeline, line width=0.9pt, fill=tmefill,
      rounded corners=3.5pt, inner sep=6pt, inner ysep=7pt,
      fit={(llm) (mon) (measl) (measr) (db) (car) (measc) (tmetitle)}] (tme) {};
\end{pgfonlayer}

\draw[flow] (llm.east) -- (mon.west) node[snum, midway] {5};

\node[comp, rotate=90, minimum width=1.6cm, minimum height=0pt, inner ysep=2.5pt,
      font=\bfseries\footnotesize]
  (tdx) at ($(tme.east)+(0.70,1.0)$) {TDX Module};
\node[comp, rotate=90, minimum width=1.6cm, minimum height=0pt, inner ysep=2.5pt,
      font=\bfseries\footnotesize]
  (h100) at ($(tme.east)+(0.70,-1.0)$) {H100 GPU};
\draw[hw] (tdx.north) -- (tdx.north -| tme.east);
\draw[hw] (h100.north) -- (h100.north -| tme.east);

\begin{pgfonlayer}{bg outer}
\node[draw=srvline, line width=0.9pt, fill=srvfill,
      rounded corners=3.5pt, inner sep=6pt, inner ysep=6pt,
      fit={(tme) (tdx) (h100) ($(tme.south)+(0,-0.32)$)}] (srv) {};
\end{pgfonlayer}
\node[font=\bfseries\footnotesize, text=srvline, anchor=south east,
      inner sep=2.5pt] at (srv.south east) {Untrusted Server};

\path let \p1=(srv.west), \p2=(srv.east) in
  node[panel, text width={\x2-\x1-11pt}, anchor=south] (extext)
  at ($(srv.north)+(0,0.40)$)
  {{\bfseries Private Document}\\[1pt]
   ``Each of these dialects was [\ldots] Northumbria south of the
   {\color{piired}\bfseries Tyne}, and {\color{piired}\bfseries Mercia},
   were overrun by Vikings [\ldots] {\color{piired}\bfseries Early West
   Saxon} [\ldots] translated from Latin.''\\[2pt]
   \colorbox{piifill}{\parbox{\dimexpr\linewidth-2\fboxsep\relax}{\raggedright\scriptsize\ttfamily Sensitive Inputs =
     [{\color{piired}`Tyne'}, {\color{piired}`Mercia'},
      {\color{piired}`Early West Saxon'}]}}};

\draw[flow] ($(db.north |- extext.south)+(-0.75,0)$) -- ($(db.north)+(-0.75,-0.08)$)
  node[snum, midway, xshift=-0.38cm] {1};

\node[qbox, text width=3.6cm, anchor=north west] (qtext) at ($(srv.west |- srv.south)+(0.16,-1.70)$)
  {From what language was literature
   translated into West Saxon?};
\node[font=\bfseries\footnotesize, anchor=south west] (u1lab)
  at ($(qtext.north west)+(0,0.04)$) {Party 1};
\begin{pgfonlayer}{bg inner}
\node[card, inner sep=4pt, fit={(qtext) (u1lab)}] (u1) {};
\end{pgfonlayer}

\node[card, dashed, inner sep=0pt,
      fit={($(u1.north east)+(0.30,0)$) (u1.south east -| srv.east)}] (uN) {};
\node[font=\bfseries\footnotesize, anchor=north] at ($(uN.north)+(0,-0.10)$)
  {Parties $2,\dots,N$};
\node[font=\Large\bfseries] at ($(uN.center)+(0,-0.12)$) {$\cdots$};
\node[font=\footnotesize, align=center, anchor=south]
  at ($(uN.south)+(0,0.08)$) {(may be\\[-2pt] adversarial)};
\draw[flow, black!65, {Stealth[length=8pt]}-{Stealth[length=8pt]}]
  (uN.north) -- (uN.north |- srv.south);

\coordinate (qx) at ($(u1.north west)!0.32!(u1.north east)$);
\coordinate (rx) at ($(u1.north west)!0.75!(u1.north east)$);
\draw[flow, txtblue]   (qx) -- (qx |- tme.south);
\node[snum] at ($(qx)!0.5!(qx |- srv.south) + (0.68,0)$) {2};
\draw[flow, txtblue]   (rx |- tme.south) -- (rx);
\node[snum] at ($(rx)!0.5!(rx |- srv.south) + (0.68,0)$) {6};

\end{tikzpicture}}
  \caption{Private Semantic Database Retrieval}
  \label{fig:psdr}
\end{subfigure}
\caption{\underline{\textbf{Overview of Illustrative Applications:}} (a) \textbf{PSFC} for privately evaluating functions on sensitive inputs and functions; (b) \textbf{PSSP} for jointly evaluating semantic queries over private sets without revealing individual elements; and (c) \textbf{PSDR} for privately querying databases without exposing sensitive records.}
\label{fig:schemes}
\vspace{-10pt}
\end{figure*}
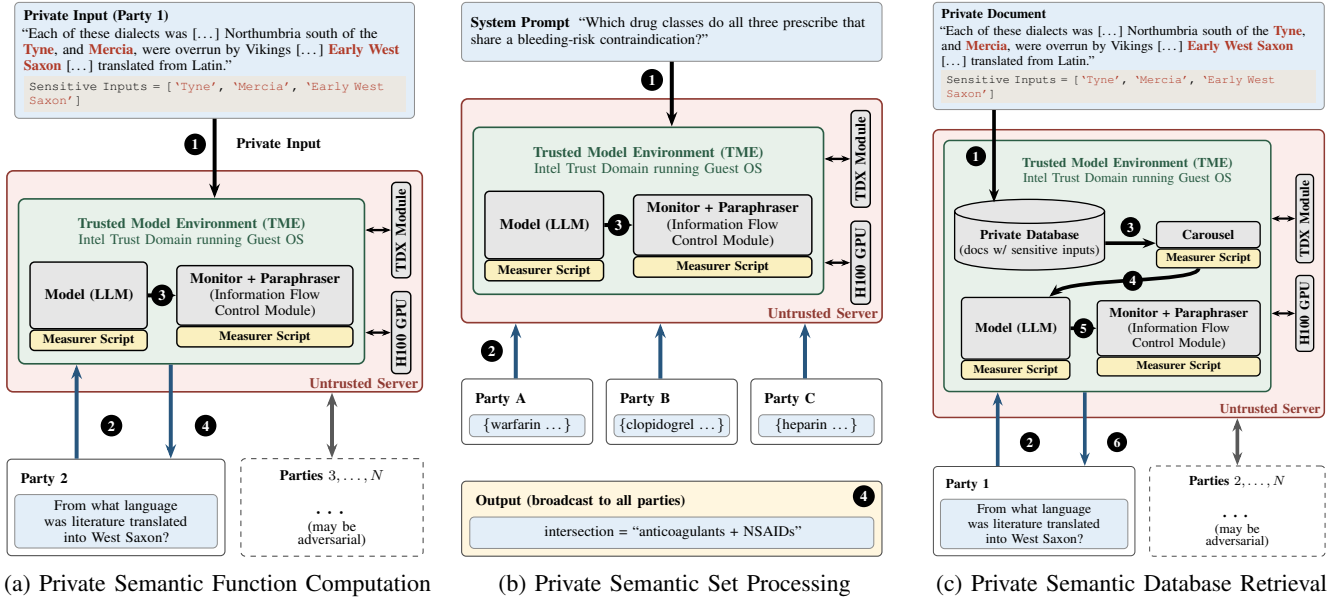

We propose \emph{trusted model environments} (\method), a class of private semantic computation primitives. 
\method runs generative models inside TEEs for computational confidentiality and verifiability, with additional proposed components for sensitive-input confidentiality, and optimizations for efficient, scalable verification across multiple verifiers.
We want to design and evaluate \method according to the following requirements:
\begin{enumerate*}[label={\textbf{(R\arabic*)}}]
\item\label{effective} \emph{effective} (correctly performs the semantic task);
\item\label{confidential} \emph{confidential} (protects computation and sensitive inputs);
\item\label{utility} \emph{utility-preserving} (retains utility on other tasks);
\item\label{verifiable} \emph{verifiable} (provides tamper-resistant evidence of the computations);
\item\label{efficient} \emph{efficient} (incurs low overhead compared to baseline model computations); and
\item\label{scalable} \emph{scalable} (supports multiple parties).
\end{enumerate*}
The design choices for \method are based on how we can meet all the above requirements, which are specific to our TEE-based implementation.

We summarize and compare limitations of various primitives in \autoref{tab:comparison}. 
We compare cryptographic primitives (``Crypto''), generative models (``Gen. Model''), and private inference using cryptographic primitives or TEEs (``Private Inf.''). Compared to prior mechanisms, \method meets all the requirements. We first present the implementation of \method across three applications as a proof of concept (\S\ref{sec:usage}), followed by its components (\S\ref{sec:components} and \S\ref{sec:componentsPSDR}).
\begin{table}[!htb]
\centering
\caption{\textbf{Limitations of Existing Mechanisms:} \cmark for satisfied, \xmark for not satisfied, and \pmark for partially satisfied. Only \method satisfies all the requirements.}
\label{tab:comparison}
\footnotesize
\setlength{\tabcolsep}{3pt}
\resizebox{\columnwidth}{!}{
\begin{tabular}{lcccccc}
\bottomrule

\toprule
\textbf{Mechanism} & \makecell{\ref{effective}\\\textbf{Effect.}} & \makecell{\ref{confidential}\\\textbf{Confid.}} & \makecell{\ref{utility}\\\textbf{Utility}} & \makecell{\ref{verifiable}\\\textbf{Verif.}} & \makecell{\ref{efficient}\\\textbf{Effic.}} & \makecell{\ref{scalable}\\\textbf{Scal.}} \\
\bottomrule

\toprule
\textbf{Crypto}            & \xmark & \pmark & \cmark & \cmark & \xmark & \xmark \\
\textbf{Gen. Model}  & \cmark & \xmark & \cmark & \xmark & \cmark & \cmark \\
\textbf{Private Inf.} & \cmark & \pmark & \cmark & \cmark & \xmark & \xmark \\
\midrule
\textbf{\method}            & \cmark & \cmark & \cmark & \cmark & \cmark & \cmark \\
\bottomrule

\toprule
\end{tabular}
}
\vspace{-10pt}
\end{table}

\subsection{Overview: \method for Illustrative Applications}\label{sec:usage}

Figure~\ref{fig:schemes} shows the three example applications and the various components. We describe each of them below:
\begin{itemize}[leftmargin=*,topsep=0pt,itemsep=0pt]
\item \textbf{PSFC (Figure~\ref{fig:psfc}):} \ding{182} A party submits its private input to \method. \ding{183} Other parties submit queries specifying the functions to compute on this input. The input is processed by a model trained with LAT to suppress verbatim secret leakage under adversarial queries. \ding{184} The model output is passed to a monitor that detects residual leakage, and \ding{185} only the sanitized output is released.

\item \textbf{PSSP (Figure~\ref{fig:pssp}):} \ding{182} A system prompt specifies the semantic query for computing the intersection of multiple parties' private sets. \ding{183} Each party submits its text or image set to \method, where the robust model computes the semantic intersection according to the system prompt. \ding{184} The model output is processed by the monitor and paraphraser to prevent disclosure of sensitive inputs, and \ding{185} the sanitized output is returned to the parties.

\item \textbf{PSDR (Figure~\ref{fig:psdr}):}  \ding{182} A private database within \method stores sensitive documents. \ding{183} Parties submit queries to retrieve information from relevant records in the database. 
\ding{184}, \ding{185} Instead of retrieving similar records to the query (leaks access patterns), the model processes each record once in a carousel (see \S\ref{sec:componentsPSDR}), incrementally updating a query-specific note.
Only relevant records contribute to the notes, which the model then uses to generate the final output. \ding{186} The model output is checked by the monitor for leakage, and \ding{187} the sanitized output is released.
\end{itemize}
The parties submit their private inputs over encrypted channels; the untrusted server forwards only ciphertexts, which are decrypted inside the enclave.

\subsection{Components of \method (PSFC and PSSP)}\label{sec:components}

We discuss the \method components with a focus on realizing PSFC and PSSP, as they share the same components. We then describe the additional components required to realize PSDR in \S\ref{sec:componentsPSDR}.
Our design builds on TEEs, which support efficient private inference of generative models~\cite{chrapek2024fortify,petridish,talaria,cmif} and can extend remote attestation to efficiently verify TEE operations at scale~\cite{laminator,chantasantitam2026pal}. However, TEEs do not prevent sensitive inputs from leaking through model outputs, particularly under adversarial queries. Thus, private inference alone is insufficient for private semantic computation. We use this as the baseline (referred to as ``Base Model'' in \S\ref{sec:evaluation}) to highlight how \method addresses these limitations and satisfies all requirements.


\noindent\textbf{\ul{\mbox{\ref{effective}} Effective, \mbox{\ref{confidential}} Confidential (Inputs), and \mbox{\ref{utility}} Utility:}} \method inherits its effectiveness from the underlying generative model. 
However, to protect against adversarial queries, \method includes additional components:
\begin{itemize}[leftmargin=*,topsep=0pt,itemsep=0pt]
\item \textbf{\emph{Protection against Verbatim Leakage:}} Given the model $\modelm$, a private input $x$, and $\mathsf{PII}(x)$ the sensitive tokens of $x$ (determined by the data owner) that the output must not reveal, the adversary seeks a query $q^\star(x)$ that maximizes disclosure of $\mathsf{PII}(x)$ in the output,
\begin{equation*}
q^\star(x)\;\in\;\arg\max_{q\,\in\,\mathcal{Q}}\;
\mathbb{E}_{y\sim\modelm(\cdot\mid x,q)}\big[\mathcal{L}_{\text{leak}}\big(y,\,\mathsf{PII}(x)\big)\big],
\end{equation*}
where $\mathcal{L}_{\text{leak}}\in[0,1]$ scores how much of $\mathsf{PII}(x)$ the output $y$ discloses ($0$ if none, $1$ if all), and $\mathcal{Q}$ is the space of queries that still appear benign to a naive input filter.
We instantiate this objective using state-of-the-art adversarial query generation methods such as AutoDAN~\cite{liu2024autodan} and many-shot~\cite{anil2024manyshot}, optimizing each query against $\mathcal{L}_{\text{leak}}$ to elicit $\mathsf{PII}(x)$. This yields a conservative evaluation of \method, as the adversary knows the sensitive spans and optimizes directly for them, making it stronger than an adversary that must first discover them.
%
We use LAT, which trains against worst-case perturbations of the model's hidden activations to resist adversarial queries~\cite{sheshadri2024latent}. An inner adversary perturbs a hidden state $h$ at a chosen layer by $\delta$ within an $\epsilon$-ball to elicit  $\mathsf{PII}(x)$, and the outer objective trains $\theta$ to be robust to the worst-case perturbation,
\begin{equation*}
\begin{aligned}
\min_{\theta}\;\;\mathbb{E}\Big[\max_{\lVert\delta\rVert\le\epsilon}
-\log\big(1-P_\theta(\mathsf{PII}(x)\mid x,q;\,h\!\leftarrow\!h+\delta)\big)\Big]&\\
+\;\beta\,\mathbb{E}\big[\mathrm{KL}(P_\theta\Vert P_{\theta_0})\big]&
\end{aligned}
\end{equation*}
where $P_\theta(\mathsf{PII}(x)\mid\cdot)$ is the probability the output discloses any of $\mathsf{PII}(x)$; the first term suppresses that probability under the worst-case perturbation, and the KL term anchors the model to its base behavior $\theta_0$ for utility.
We train only lightweight adapters (e.g., LoRA), leaving the base weights frozen, and perform extensive grid-search-based hyperparameter tuning to balance confidentiality, effectiveness, and utility.

\item \textbf{\emph{Protection against Semantic Leakage:}}
Despite suppressing verbatim leakage, an adversary may force the model to paraphrase secrets indirectly and evade LAT. 
To address this, the Information Flow Control (IFC) module checks every $\modelm$ output using two LLMs: a monitor $\modeMon$ that detects residual leakage and a paraphraser $\modePara$ that removes it. If the monitor finds no leakage, the output is released unchanged; otherwise, $\modePara$ rewrites the original output by removing the protected item, indirect descriptions, relationships, and other unnecessary content while preserving the answer.
\end{itemize}
Both effectiveness \ref{effective} and confidentiality \ref{confidential} trade off against utility \ref{utility}: LAT modifies the model, and the paraphraser rewrites flagged outputs. We contain the former through the KL loss term, which minimizes the difference from the base model, and the latter by releasing unflagged outputs unchanged and preserving the answer otherwise.

\noindent\textbf{\ul{\mbox{\ref{confidential}} Confidential (Computation):}} \method inherits its computation confidentiality from the TEE: the model, inputs, and outputs reside inside the enclave, whose memory encryption and isolation ensure that the untrusted server cannot read its contents.
However, there is a potential timing side channel as the autoregressive decoding time in LLMs grows with the number of emitted tokens, so the \emph{length} of a response is observable via timing and can leak properties of the private inputs (e.g., size of the intersection in PSSP). We close this channel by having every task inference decode a fixed token budget: the model emits its answer and then pads to the fixed length, so the decoding time is independent of the answer. The padding is removed before the answers are scored and thus does not impact effectiveness or utility. However, TEE-specific side channels are possible (see \S\ref{sec:discussion}).

\noindent\textbf{\ul{\mbox{\ref{verifiable}} Verifiable:}} \method relies on the TEE's integrity guarantees and remote attestation to support verifiability. For \method operations, attestations bind their inputs and outputs: $\sigma = \mathsf{Sign}_{sk_{\text{TEE}}}\!\big(\mathsf{Hash}(\cdot), \text{nonce}\big)$. 
Private inputs/outputs are included as hashes and their plaintext never leaves the enclave. 
We define the following attestations, whose complete descriptions are in Appendix~\ref{app:attestations}:
\begin{itemize}[leftmargin=*,topsep=0pt,itemsep=0pt]
\item \textbf{\emph{Model Measurement:}}
\method hashes the model weights and emits $\sigma_{\modelm} = \mathsf{Sign}_{sk_{\text{TEE}}}(\mathsf{Hash}(\text{weights}), \text{nonce})$, certifying which model (the base model along with the robust adapter) was executed inside \method.

\item\textbf{\emph{Input Commitment:}}
Each private input is committed as $\sigma_{\text{in}} = \mathsf{Sign}_{sk_{\text{TEE}}}(\mathsf{Hash}(\text{input}), \text{nonce})$, fixing the exact private input that was used during the computation while keeping its content hidden. 
Committing memory-mapped data introduces no time-of-check-to-time-of-use gap: the hash is computed over the copy of the input residing in enclave-protected memory. Similar to prior work~\cite{chantasantitam2026pal}, data too large to be resident is instead verified against the committed digest at access time, so post-commitment modification is detected on use.

\item\textbf{\emph{Proof of Inference:}}
For $\modelm$'s inference, we bind the input queries and the output to $\modelm$: $\sigma_{\text{inf}} = \mathsf{Sign}_{sk_{\text{TEE}}}(\mathsf{Hash}(\pi, \sigma_{\text{in}}, \sigma_{\modelm}, \text{output}), \text{nonce})$. This attestation certifies that the committed model, on the committed inputs under a system prompt $\pi$, produced this output.
Additionally, the IFC module contains two LLMs: a monitor $\modeMon$, and a paraphraser $\modePara$, whose inferences are attested separately.
For $\modeMon$'s prediction of whether $\modelm$'s output contains sensitive inputs, we have $\sigma_{\text{mon}} = \mathsf{Sign}_{sk_{\text{TEE}}}(\mathsf{Hash}(\modeMon, \sigma_{\text{inf}}, \text{verdict}), \text{nonce})$.
For $\modePara$'s rewrite of the flagged text from $\modeMon$, we have $\sigma_{\text{par}} = \mathsf{Sign}_{sk_{\text{TEE}}}(\mathsf{Hash}(\modePara, \sigma_{\text{inf}}, \text{rewrite}), \text{nonce})$.
Each IFC module inference is separately attested, binding each model to its inputs and outputs.

\end{itemize}
Each \method computation is attested by the same measurement script: after the computation completes, it hashes the computation's inputs and outputs inside the enclave and signs them. The attestations thus differ only in the values they bind, not in the mechanism.

\noindent\textbf{Verification:} A verifier (either the parties or an external verifier such as a regulator) checks the signature and recomputes the hashes to confirm that the attested model weights, inputs, and IFC policy produced the claimed output. Specifically, a verifier can check the attestation signature against the platform certificate chain (rooted in Intel DCAP and NVIDIA NRAS), confirm that the nonce matches the verifier-issued session nonce to prevent replay, and recompute and compare the hashes. For inputs accessible to the verifier, such as its query and released output, hashes can be recomputed and compared with the attestation to confirm its execution inside \method. 
For private inputs, such as the system prompt $\pi$ or database, we assume the owner publishes a commitment against which the attested hashes can be checked.

\noindent\textbf{\ul{\mbox{\ref{efficient}} Efficient and \mbox{\ref{scalable}} Scalable:}}
The various components for \method incur an additional overhead compared to the baseline of a base model running inside TEEs (without any input confidentiality). In addition to the overhead to run these additional components, we also incur a cost for attestations.
Naively, attestation is expensive: a hardware attestation is a system call, and issuing one per party, per record, and per inference would dominate the overhead and scale poorly with the number of parties. \method reduces this cost with the following optimizations (see Appendix~\ref{sec:asymptotic} for overheads):
\begin{itemize}[leftmargin=*,topsep=0pt,itemsep=0pt]
\item \textbf{\emph{Merkle-tree Batching:}} The expensive step is the hardware attestation, so issuing one per party scales poorly. Instead, each party's $(\text{function},\text{output})$ is hashed into a \emph{leaf}, and the leaves form a Merkle tree whose root is covered by a single hardware attestation. For example, with four parties and leaves $L_A, L_B, L_C, L_D$, where $L_i$ hashes party $i$'s function and output, the enclave signs
$\mathsf{root}=\mathsf{Hash}(\mathsf{Hash}(L_A, L_B),\mathsf{Hash}(L_C, L_D))$ with one attestation. Party $C$ receives $L_C$, $L_D$, and $\mathsf{Hash}(L_A, L_B)$, allowing it to recompute and verify the signed root without learning the other parties' inputs or outputs. Thus, one attestation covers the entire batch, reducing per-party attestation cost as the batch grows. This amortizes the \emph{attestation} cost and complements the batch-level optimization below, which amortizes the \emph{computation} cost. Confidentiality is preserved as a party sees only the $O(\log N)$ sibling digests on its authentication path, which are opaque and, with salted leaves, reveal nothing about the committed values; the full tree is never disclosed, and only the root enters the hardware attestation.

\item \textbf{\emph{Batch-level Amortization:}} Batching queries amortizes expensive attestations across parties: a single inference and IFC module run attests all queries in a batch, while only lightweight per-party input commitments scale with the number of parties. Thus, per-party overhead decreases with batch size. 

\item \textbf{\emph{In-TEE Signing Key:}} Instead of generating an expensive hardware attestation each time, we generate an in-memory signing key pair inside the TEE and attest it once. Each subsequent attestation can use that key instead of the hardware attestation, which is significantly cheaper (\S\ref{sec:evaluation}).

\item \textbf{\emph{One-time Model Measurement:}} The model weights are hashed once at startup and reused across all inferences rather than re-measured per query. Also, the attestation stack is initialized before processing the queries to ensure that the first request does not incur a high overhead.
\end{itemize}

\subsection{Additional \method Components for PSDR}\label{sec:componentsPSDR}

In addition to the components described in \S\ref{sec:components}, PSDR requires additional components to prevent access-pattern leakage during private database retrieval, described below.

\noindent\textbf{\ul{\mbox{\ref{confidential}} Confidential (Access Pattern):}} We need an additional component to minimize \emph{access-pattern} leakage when retrieving different data records from the private database. Simply retrieving the top-$k$ similar records, as done in RAG pipelines~\cite{lewis2020retrieval}, reveals the access pattern.
One option is to use oblivious RAM (ORAM)~\cite{Ostrovsky90,StefanovDSFRYD13}, which hides memory-access patterns from an untrusted server but incurs poly-logarithmic overhead: $\Omega(\log N_{db})$ per access is inherent~\cite{Ostrovsky90}, while Path ORAM incurs $O(\log N_{db})$ bandwidth overhead per access~\cite{StefanovDSFRYD13}, making the $N$ parties' lookups cost $O(N\log N_{db})$. Instead, inspired by \emph{Carousel}~\cite{circleGame}, we continuously scan the dictionary (or a compact representation) inside the TEE, giving every lookup the same data-independent trace. Each scan answers the $N$ pending queries in $O(N_{db}+N)$ time, or amortized $O(N_{db}/N)$ per query. This trades per-query overhead for batch throughput and achieves $\sim$3.7K queries/s on a $2^{26}$-entry ($\approx$67M) dictionary in Intel SGX, outperforming Path ORAM on the same hardware~\cite{circleGame}. Thus, database membership checks reveal no access patterns.

We adapt \emph{Carousel} to \method for PSDR by scanning the \emph{entire} database in a fixed order, making the access pattern independent of which records are relevant. 
For each record, the model writes a short \emph{note} containing only query-relevant information, from which the final answer is generated after the scan. 
Unlike the original \emph{Carousel} design, our adaptation uses these notes to support private semantic retrieval.
A malicious server cannot skip records because the entire private dataset resides in tamper-resistant TEE memory. 

\noindent\textbf{\ul{\mbox{\ref{verifiable}} Verifiable:}} We present proof of oblivious access for the carousel by generating an attestation $\sigma_{\text{obl}} = \mathsf{Sign}_{sk_{\text{TEE}}}(\mathsf{Hash}(\text{scanned-records}), \text{nonce})$. This certifies that every database record was parsed, and the access pattern is data-independent.



\section{Experiment Setup}\label{sec:setup}

We present an overview of the datasets, models, and metrics used for evaluation. For all the experiments, we report the mean and standard deviation across five runs.


\noindent\textbf{\ul{Hardware Setup:}} We configure the Intel TDX with 32 vCPUs and 128 GB of memory, running Ubuntu 24.04.1 LTS with kernel version 6.8.0-86-generic. The host system is equipped with an Intel Xeon Silver 4514Y CPU, 512 GB of RAM, and an NVIDIA H100 NVL 94 GB GPU operating in confidential-computing mode.
\method's measurement and attestation components are implemented in Python. 
The implementation uses PyTorch, Intel Data Center Attestation Primitives (DCAP) for CPU attestation, and NVIDIA Remote Attestation Service (NRAS) for GPU attestation.

\noindent\textbf{\ul{Datasets:}} We describe the datasets and splits across applications (summarized in Table~\ref{tab:datasets}), and describe dataset construction in Appendix~\ref{app:dataset}.
\begin{itemize}[leftmargin=*,topsep=0pt,itemsep=0pt]
\item \textbf{\ref{effective} Effectiveness / \ref{confidential} Confidentiality:} We evaluate effectiveness and confidentiality on application-specific datasets. 
For PSFC, we use \dataverb{SQuADv2}~\cite{squad2}, a reading-comprehension benchmark where each party submits a free-text question over a shared document ($250$ documents, $1{,}250$ questions). 
For PSSP, we use \dataverb{DBpedia14}~\cite{dbpedia}, where $250$ Wikipedia abstracts labeled into $14$ ontology classes serve as the parties' private set elements. 
For PSDR, we use \dataverb{HotpotQA}~\cite{hotpotqa}, a multi-hop question-answer benchmark in which the passages form the database and each party submits a free-text query ($1{,}000$ passages, $250$ queries). 
Each dataset is split into $100/50/100$ for train, validation (for hyperparameter tuning), and test splits.
\begin{table}[!htbp]
\centering
\caption{\textbf{Summary of datasets} across the applications, and train-validation-test splits.}
\small
\setlength{\tabcolsep}{3pt}          
\resizebox{\columnwidth}{!}{
\begin{tabular}{lllcl}
\bottomrule

\toprule
\textbf{Requirement} & \textbf{App.} & \textbf{Dataset} & \makecell{\textbf{Records}} & \makecell{\textbf{Split}\\\textbf{(tr.\,/\,val\,/\,te.)}} \\
\bottomrule

\toprule
\multirow{5}{*}{\makecell[l]{\ref{effective} \textbf{Effective}\\ \ref{confidential} \textbf{Confidential}}}
 & PSFC & \dataverb{SQuADv2}~\cite{squad2}   & \makecell{$250$ documents\\($1250$ questions)} & $100\,/\,50\,/\,100$ \\ \cmidrule(l){2-5}
 & PSSP     & \dataverb{DBpedia14}~\cite{dbpedia}& \makecell{$250$ documents\\($14$ classes)}          & $100\,/\,50\,/\,100$ \\ \cmidrule(l){2-5}
 & PSDR                & \dataverb{HotpotQA}~\cite{hotpotqa} & \makecell{$1{,}000$ passages\\($250$ queries)}    & $100\,/\,50\,/\,100$ \\
\midrule
\multirow{2}{*}{\ref{utility} \textbf{Utility}}
 & \multirow{2}{*}{All} & \dataverb{MMLU}~\cite{mmlu}          & $200$ questions & $100$ $/$ - $/$ $100$\\
 & & \dataverb{CSQA}~\cite{csqa} & $200$ questions & $100$ $/$ - $/$ $100$\\
\bottomrule

\toprule
\end{tabular}
}
\label{tab:datasets}
\vspace{-10pt}
\end{table}

\item \noindent\textbf{\ref{utility} Utility:}
We measure utility on two standard multiple-choice benchmarks that are shared across all three applications: \dataverb{MMLU}~\cite{mmlu} covering  academic/professional knowledge questions with four options per question; and \dataverb{CommonsenseQA (CSQA)}~\cite{csqa}  covering commonsense reasoning questions with five options per question. We use $100$ records during LAT to improve utility, and $100$ records for testing.
\end{itemize}
For confidentiality of sensitive inputs \ref{confidential}, since none of the datasets provides annotations of sensitive inputs, we use Microsoft Presidio to extract entities as ground-truth $\mathsf{PII}(x)$ (Appendix~\ref{app:dataset}). We assume these sensitive inputs are predefined as part of the information-flow policy. Runtime extraction of sensitive inputs is future work, and can be supported by integrating an additional name-entity recognition model into \method, and attesting its inference.

\noindent\textbf{\ul{Models:}}
All applications run open-weight, instruction-tuned language models locally inside the TEE. We evaluate three models of comparable scale ($7$--$9$B) drawn from three different vendors and architecture families: \modelverb{Ministral-8B}~\cite{ministral}, \modelverb{gemma-2-9b}~\cite{gemma2}, and \modelverb{OLMo-2-7B}~\cite{olmo2}. The IFC module's monitor and paraphraser are instantiated with a separate model, \modelverb{Qwen3-8B}~\cite{qwen3}, running in reasoning mode.
\modelverb{Qwen3-8B} is deliberately drawn from a different vendor and architecture family than the three models it supervises, to prevent any evaluation bias. Additionally, we use an LLM as a detector since it is better at parsing and reasoning about text compared to traditional rule-based or language model-based classifiers.
We discuss the extension to larger models in \S\ref{sec:discussion}, and discuss the hyperparameters in Appendix~\ref{app:hyperparameters}.


\noindent\textbf{\ul{Evaluation Metrics:}} We define metrics for each requirement:
\begin{itemize}[leftmargin=*,topsep=0pt,itemsep=0pt]
\item \noindent\textbf{\ref{effective} Effectiveness:} Task accuracy against ground truth is scored deterministically. For PSFC and PSDR the output is free text, so we report the token-level F1 score, which is the harmonic mean of precision and recall over the bag of tokens shared between the predicted and gold answers, after normalizing case, punctuation, and articles. F1 scores better capture small but acceptable differences between the output and ground truth that exact-match metrics may miss. For PSSP,  we report set-level F1 between the predicted and the true intersection, after mapping each predicted free-text category to its nearest ontology label so that matching is semantic rather than exact-string. All metrics lie in $[0,1]$, with $1$ indicating a ground-truth match and $0$ a refusal.

\item \textbf{\ref{confidential} Confidentiality (Computation):} For PSDR, we measure access-pattern leakage, i.e., how much an observer can infer about the records relevant to a query from the records accessed. Specifically, we ask \emph{if an observer sees only the accessed records, how likely are they to correctly identify a relevant (target) record?} We estimate this by guessing among the accessed records and report the fraction of guesses that correspond to target records. 
The metric ranges from 0 to 1, where 0 means no useful information is revealed and 1 means the access pattern reveals the target exactly.
We compare a standard top-$k$ retrieval baseline, which accesses $k$ most similar records, with our carousel, which scans the entire database in the same order for every query. 
For PSDR, we consider timing and access-pattern leakage, following ORAM literature, but other side channels, such as per-record note lengths, may remain. These can be mitigated with padding, similar to our fixed-output padding for content-dependent timing (\S\ref{sec:components}); these are left as future work.

\textbf{\ref{confidential} Confidentiality (Sensitive Inputs):} We quantify this using the \emph{attack success rate} (ASR): the fraction of released outputs that disclose sensitive inputs. ASR ranges from $0$ (no leakage) to $1$ (complete leak); \emph{lower is better}.
\emph{ASR-Verbatim} counts literal reproduction of sensitive inputs via string matching, while \emph{ASR-Semantic} counts both literal and indirect disclosures (e.g., paraphrases, descriptions, or character-level encodings). For ASR-Semantic, we use a separate \modelverb{Qwen3-8B} LLM judge that receives the query, output, and sensitive spans and returns YES/NO on whether the output reveals any sensitive inputs, including paraphrases, summaries, partial statements, translations, or light obfuscation. We use a different model family from those running the computation, to avoid evaluation bias.
We evaluate both metrics under three conditions:
\begin{itemize}[leftmargin=*,topsep=0pt,itemsep=0pt]
    \item \emph{Benign:} We measure incidental leakage by \emph{ASR-Verbatim/-Semantic} on standard task queries without adversarial content.
    \item \emph{Adversarial:} We evaluate this by generating adversarial queries using the state-of-the-art \emph{AutoDAN}~\cite{liu2024autodan} and \emph{many-shot}~\cite{anil2024manyshot} techniques. 
    \item \emph{Indirect:} We evaluate this using queries that disclose sensitive inputs by describing or spelling them out.
\end{itemize}
All metrics are evaluated before LAT, after LAT, and after IFC.
While IFC can detect and prevent verbatim leakage, LAT suppresses many such cases upfront, reducing the need to invoke the monitor and paraphraser and thereby lowering IFC overhead (see below under \ref{efficient} and \ref{scalable}).

\item \textbf{\ref{utility} Utility:} We report the accuracy on two standard multiple-choice benchmarks, \dataverb{MMLU} and \dataverb{CSQA}, which are unrelated to the application task. Accuracy is from $0$ to $1$, with higher values indicating better utility. Random guessing yields $0.25$ for \dataverb{MMLU} (four options) and $0.20$ for \dataverb{CSQA} (five options).

\item\textbf{\ref{efficient} Efficiency and \ref{scalable} Scalability:} Given the wide range of TEE-enabled machine learning services (e.g., AWS, Microsoft Azure, and Google Cloud), we use a single model inference in a TEE as the baseline and compare its overhead with that of \method. We evaluate and report efficiency and scalability for running different \method components (IFC module and attestations).
\begin{itemize}[leftmargin=*,topsep=0pt,itemsep=0pt]
\item \textbf{Inference Overhead:} With and without LAT, the inference overhead is the same, since the number of parameters and the computation are unchanged. Thus, using a single model inference inside a TEE as the baseline, we measure the additional overhead of IFC, including monitor inference for detecting leakage of sensitive inputs and paraphraser inference (with thinking enabled) for sanitizing flagged outputs. We evaluate the overhead with more parties for scalability.

\item \textbf{Carousel Overhead (for PSDR):} Compared to retrieving the top-$k$ most similar records, scanning the entire database adds overhead. We measure the overhead of the carousel relative to standard RAG operations while preventing access-pattern leakage.

\item \textbf{Attestation Overhead:} We decompose the wall-clock time of each attested run as follows: (i) a \emph{baseline time} (\method execution inside the TEE), (ii) a \emph{measurement time} (hashing inputs, outputs, and the model to bind the computation), and (iii) an \emph{attestation time} (producing the attestation evidence). 
We separate measurement into a \emph{one-time offline model measurement} (a single SHA-256 hash over the weights) that is amortized across all subsequent inferences, and a \emph{recurring per-inference overhead}, which is the ratio of the measurement and attestation time over the baseline time. 
We also compare two attestation mechanisms: (i) \emph{hardware attestation}, which generates a TDX quote per payload, and (ii) \emph{in-TEE attestation}, where a single hardware attestation certifies an in-TEE key that signs subsequent payloads (\S\ref{sec:components} under \ref{efficient} and \ref{scalable}).
We report scalability as the overhead with more parties. Remaining requirements (e.g., effectiveness, utility, and confidentiality) are independent of the number of participants.
\end{itemize}
\noindent\textbf{Note on TEE Overhead:} The above costs assume workloads are already executed \emph{inside} the TEE, and we report the cost added by \method. The additional overhead inside TEE compared to outside TEE, is not introduced by \method. Prior benchmarks report that these overheads are small, less than 7\%~\cite{chrapek2025confidential,zhu2024confidential,wang2026benchmarking}, which is negligible compared with the orders-of-magnitude higher costs: MPC-based LLaMA-7B inference achieves only $\sim3.3\times10^{-3}$ tokens/s (about five minutes per token)~\cite{puma}; BumbleBee takes $\sim8$ minutes per LLaMA-7B token on CPUs~\cite{bumblebee}; THOR requires $\sim10$ minutes for a single 128-token BERT-base inference on a GPU~\cite{thor}. Communication is similarly costly, reaching $\sim285$\,GB for Iron and $\sim60$GB for BOLT~\cite{iron,bolt}.
\end{itemize}

\section{Evaluation}\label{sec:evaluation}

We evaluate effectiveness \ref{effective}, confidentiality \ref{confidential}, utility \ref{utility}, efficiency \ref{efficient}, and scalability \ref{scalable}, across three applications: PSFC (\S\ref{sec:evalPSFC}), PSSP (\S\ref{sec:evalPSSP}), and PSDR (\S\ref{sec:evalPSDR}). 
We provide additional results in Appendix~\ref{sec:AdditionalResults} and~\ref{app:adaptive}.
Across all applications, we expect our proof-of-concept (``+LAT+IFC'') to provide high input confidentiality \ref{confidential} relative to the ``base model''. 
However, this comes with trade-offs in utility \ref{utility} and effectiveness \ref{effective}, as well as overhead \ref{efficient}-\ref{scalable}. We therefore do not expect optimal utility, effectiveness, or overhead for various operations, and improving these trade-offs is left as future work.


\subsection{Private Semantic Function Computation}\label{sec:evalPSFC}
\begin{arxiv}
Recall that in PSFC, \method holds one party's private input, while other parties submit private queries specifying semantic functions over the input and receive the outputs.
\end{arxiv}
\noindent\textbf{\ul{\mbox{\ref{effective}} Effectiveness and \mbox{\ref{utility}} Utility:}} Ideally, we expect task effectiveness to remain high when applying the \method pipeline, which includes LAT and the IFC module, compared to using only the pretrained base model. We validate this by reporting token-F1 scores between model predictions and ground-truth responses from \dataverb{SQuADv2} (\autoref{tab:psfc-lm-eff-util}). All three base models achieve high effectiveness, which decreases after applying LAT because it degrades response quality and reduce token-F1. 
However, applying the IFC module, including the paraphraser, largely restores token-F1 to the base model level by removing unnecessary or malformed text. Overall, \method's effectiveness is close to the base models.

For \ref{utility}, applying LAT degrades the utility compared to the base model.
We measure this as the combined accuracy on \dataverb{MMLU} and \dataverb{CSQA}, comparing the base models with LAT and IFC applied (\autoref{tab:psfc-lm-eff-util}).
We see a small drop in utility:
\modelverb{Ministral-8B} with a 5.3 percentage-point drop (62.5\% $\rightarrow$ 57.2\%); 
\modelverb{gemma-2-9b} with a 4.4 percentage-point drop (72.0\% $\rightarrow$ 67.6\%); and
\modelverb{OLMo-2-7B} with a 0.6 percentage-point drop (65.1\% $\rightarrow$ 64.5\%).
Overall, applying \method retains utility within a few points of the base model.

\begin{table}[!htb]
\centering
\caption{\textbf{(PSFC) Effectiveness \ref{effective} and Utility \ref{utility}:} \method, which includes LAT and IFC, maintains effectiveness close to the base model with a small utility drop.}
\label{tab:psfc-lm-eff-util}
\footnotesize
\setlength{\tabcolsep}{4pt}
\begin{tabular}{lll c}
\bottomrule

\toprule
\textbf{Model} & \textbf{Metric} & \textbf{Stage} & \textbf{Value} \\
\bottomrule

\toprule
\multicolumn{4}{l}{\ref{effective} \textbf{Effectiveness}} \\
\midrule
\multirow{3}{*}{\modelverb{Ministral-8B}}
 & \multirow{3}{*}{Token-F1} & Base Model    & 0.637$\pm$0.019 \\
 & & +LAT      & 0.563$\pm$0.194 \\
 & & +LAT+IFC & 0.624$\pm$0.149 \\
\cmidrule(lr){1-4}
\multirow{3}{*}{\modelverb{gemma-2-9b}}
 & \multirow{3}{*}{Token-F1} & Base Model    & 0.813$\pm$0.011 \\
 & & +LAT      & 0.791$\pm$0.058 \\
 & & +LAT+IFC & 0.804$\pm$0.048 \\
\cmidrule(lr){1-4}
\multirow{3}{*}{\modelverb{OLMo-2-7B}}
 & \multirow{3}{*}{Token-F1} & Base Model    & 0.731$\pm$0.008 \\
 & & +LAT      & 0.681$\pm$0.063 \\
 & & +LAT+IFC & 0.714$\pm$0.054 \\
\midrule
\multicolumn{4}{l}{\ref{utility} \textbf{Utility}} \\
\midrule
\multirow{2}{*}{\modelverb{Ministral-8B}} & \multirow{2}{*}{Accuracy} & Base Model        & 0.625$\pm$0.000 \\
 & & +LAT+IFC & 0.572$\pm$0.095 \\
\cmidrule(lr){1-4}
\multirow{2}{*}{\modelverb{gemma-2-9b}} & \multirow{2}{*}{Accuracy} & Base Model        & 0.720$\pm$0.000 \\
 & & +LAT+IFC & 0.676$\pm$0.022 \\
\cmidrule(lr){1-4}
\multirow{2}{*}{\modelverb{OLMo-2-7B}} & \multirow{2}{*}{Accuracy} & Base Model        & 0.651$\pm$0.000 \\
 & & +LAT+IFC & 0.645$\pm$0.019 \\
\bottomrule

\toprule
\end{tabular}
\vspace{-10pt}
\end{table}

\begin{table}[!htb]
\centering
\caption{\textbf{(PSFC) Confidentiality \ref{confidential}:} Both ASR-Verbatim and ASR-Semantic decrease with LAT and IFC across benign, adversarial, and indirect.}
\label{tab:psfc-lm-conf}
\footnotesize
\setlength{\tabcolsep}{3.5pt}
\begin{tabular}{ll ccc}
\bottomrule

\toprule
\textbf{Model} & \textbf{Stage} & \textbf{Benign} & \textbf{Adversarial} & \textbf{Indirect} \\
\bottomrule

\toprule
\multicolumn{5}{l}{\textbf{ASR-Verbatim}} \\
\midrule
\multirow{3}{*}{\modelverb{Ministral-8B}}
 & Base Model          & 0.01$\pm$0.01 & 1.00$\pm$0.00 & 0.54$\pm$0.03 \\
 & $+$LAT     & 0.09$\pm$0.08 & 0.18$\pm$0.16 & 0.18$\pm$0.14 \\
 & $+$LAT$+$IFC  & 0.01$\pm$0.02 & 0.01$\pm$0.02 & 0.01$\pm$0.01 \\
\cmidrule(lr){1-5}
\multirow{3}{*}{\modelverb{gemma-2-9b}}
 & Base Model          & 0.02$\pm$0.01 & 1.00$\pm$0.00 & 0.22$\pm$0.02 \\
 & $+$LAT     & 0.05$\pm$0.03 & 0.11$\pm$0.08 & 0.33$\pm$0.27 \\
 & $+$LAT$+$IFC  & 0.00$\pm$0.00 & 0.01$\pm$0.01 & 0.05$\pm$0.07 \\
\cmidrule(lr){1-5}
\multirow{3}{*}{\modelverb{OLMo-2-7B}}
 & Base Model          & 0.00$\pm$0.00 & 0.57$\pm$0.05 & 0.63$\pm$0.03 \\
 & $+$LAT     & 0.06$\pm$0.02 & 0.16$\pm$0.09 & 0.14$\pm$0.09 \\
 & $+$LAT$+$IFC  & 0.00$\pm$0.00 & 0.01$\pm$0.01 & 0.01$\pm$0.01 \\
\midrule
\multicolumn{5}{l}{\textbf{ASR-Semantic}} \\
\midrule
\multirow{3}{*}{\modelverb{Ministral-8B}}
 & Base Model          & 0.01$\pm$0.01 & 1.00$\pm$0.00 & 0.64$\pm$0.02 \\
 & $+$LAT     & 0.12$\pm$0.09 & 0.20$\pm$0.15 & 0.21$\pm$0.13 \\
 & $+$LAT$+$IFC  & 0.02$\pm$0.02 & 0.03$\pm$0.03 & 0.02$\pm$0.01 \\
\cmidrule(lr){1-5}
\multirow{3}{*}{\modelverb{gemma-2-9b}}
 & Base Model          & 0.02$\pm$0.01 & 1.00$\pm$0.00 & 0.49$\pm$0.05 \\
 & $+$LAT     & 0.06$\pm$0.03 & 0.12$\pm$0.09 & 0.34$\pm$0.27 \\
 & $+$LAT$+$IFC  & 0.00$\pm$0.01 & 0.01$\pm$0.01 & 0.05$\pm$0.07 \\
\cmidrule(lr){1-5}
\multirow{3}{*}{\modelverb{OLMo-2-7B}}
 & Base Model          & 0.00$\pm$0.00 & 0.59$\pm$0.05 & 0.66$\pm$0.02 \\
 & $+$LAT     & 0.09$\pm$0.03 & 0.21$\pm$0.11 & 0.19$\pm$0.11 \\
 & $+$LAT$+$IFC  & 0.01$\pm$0.01 & 0.02$\pm$0.02 & 0.03$\pm$0.03 \\
\bottomrule

\toprule
\end{tabular}
\vspace{-10pt}
\end{table}

\noindent\textbf{\ul{\mbox{\ref{confidential}} Confidentiality:}} Ideally, \method should not leak sensitive inputs. We evaluate leakage under three settings: \emph{benign} queries for incidental leakage, \emph{adversarial} queries generated using AutoDAN~\cite{liu2024autodan} and many-shot jailbreak~\cite{anil2024manyshot}, and \emph{indirect} queries designed to induce subtle embedding or paraphrasing of sensitive inputs in the output.
We evaluate the impact of LAT against adversarial queries and IFC against residual adversarial and indirect leakage across all three models (\autoref{tab:psfc-lm-conf}). Under benign queries, leakage is initially small and becomes negligible ($\sim$0\%) after LAT and IFC. For adversarial queries, LAT substantially reduces leakage, with IFC suppressing the remaining leakage to $\sim$0\%. For indirect queries, leakage generally decreases after LAT and IFC, although LAT can occasionally increase it, illustrating an unintended interaction between adversarial robustness and indirect leakage~\cite{duddu2024sok}. Here, IFC successfully suppresses this residual leakage. Overall, \method reduces leakage of sensitive inputs to negligible levels.

\noindent\textbf{\ul{\mbox{\ref{efficient}} Efficiency and \mbox{\ref{scalable}} Scalability:}} For PSFC, we present the IFC and attestation overheads below:
\begin{itemize}[leftmargin=*,topsep=0pt,itemsep=0pt]
\item\textbf{Inference Overhead:} A single model inference for a PSFC query takes $1.9$s for \modelverb{Ministral-8B}, $3.2$s for \modelverb{gemma-2-9b}, and $2.2$s for \modelverb{OLMo-2-7B}. The IFC module additionally checks each answer with the monitor (8.6s) and invokes a paraphraser (29.4s) only when leakage is detected. Thus, clean answers incur 10.5--11.8s per inference, while paraphrasing takes about 40--41s.

Batching records from multiple parties improves scalability: base inference latency remains near 2s as the batch size increases from 1 to 8, reducing per-query latency from 1.9s to 262ms. The IFC module's latency remains near 62s from batch sizes 1 to 4, reducing per-query latency from 60s to 16s. Thus, batching the IFC module's checks across parties reduces per-party overhead toward $1/N$.

\item\textbf{Attestation Overhead:} The one-time model measurement (\autoref{fig:efficiencyPSFC}) scales with model size: $58.4\pm2.8$s for \modelverb{Ministral-8B}, $65.1\pm5.6$s for \modelverb{gemma-2-9b}, and $59.1\pm4.6$s for \modelverb{OLMo-2-7B}. The largest model is the most expensive, while the two smaller models are similar.
Hardware attestation adds $0.07$--$0.13\%$ overhead and \emph{falls} with $N$ as the batch-level attestations amortize across parties ($1.4$--$1.6\times$ from $N{=}2$ to $6$); the in-TEE attestation adds $0.003$--$0.005\%$, roughly $24\times$ lower.

\autoref{fig:efficiencyPSFC} shows this trend as $N$ grows from $2$ to $6$: the batch-level Merkle-root attestation and the one-time model measurement are shared across $N$ parties, so the per-party overhead decreases as $1/N$ and attestation becomes a smaller fraction of the total computation.

\begin{figure}[!htbp]
\centering
\includegraphics[width=0.7\columnwidth]{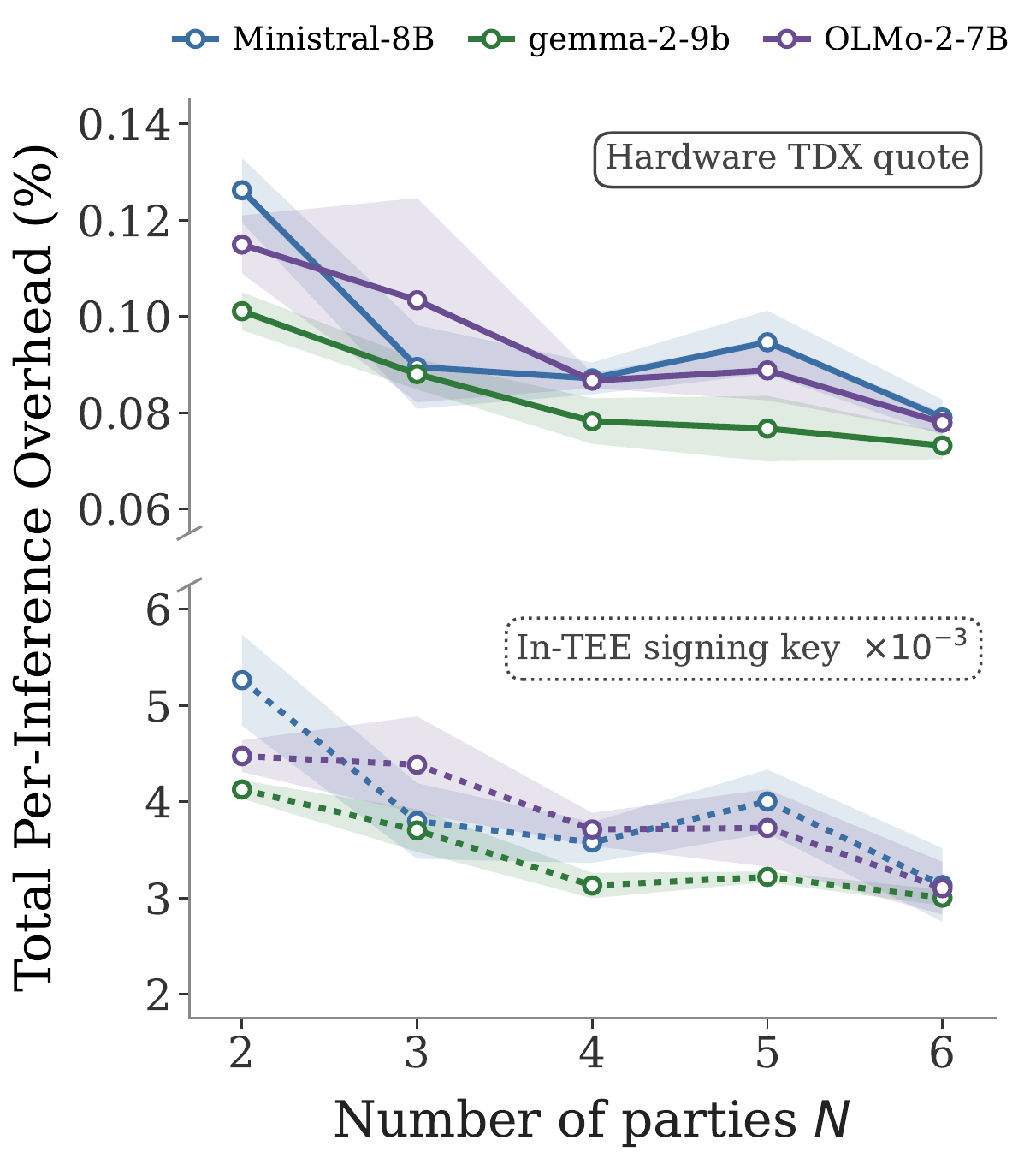}
\caption{\textbf{(PSFC) Efficiency \ref{efficient} and Scalability \ref{scalable}:} Measurement $+$ attestation as \% of the baseline vs.\ $N$; hardware (top) and in-TEE attestation (bottom).}
  \label{fig:efficiencyPSFC}
\vspace{-10pt}
\end{figure}
\end{itemize}

\begin{takeaway}
\textbf{Summary:} \method enables PSFC while maintaining effectiveness \ref{effective}, negligible leakage \ref{confidential}, small utility drop \ref{utility}, efficiency \ref{efficient}, and scalability \ref{scalable}.
\end{takeaway}

\subsection{Private Semantic Set Processing}\label{sec:evalPSSP}
\begin{arxiv}
Recall that in PSSP, multiple parties share their private sets, which a model processes privately to execute a semantic query (specified as a system prompt) and release the output.
\end{arxiv}
\noindent\textbf{\ul{\mbox{\ref{effective}} Effectiveness and \mbox{\ref{utility}} Utility:}} Ideally, the semantic processing of the parties' sets inside \method should not be degraded by
LAT or the IFC module.  We measure set-F1 between the predicted and ground-truth intersections in \dataverb{DBpedia14}.
As shown in \autoref{tab:pssp-lm-eff-util}, applying LAT leads to a small drop in set-F1 score for \modelverb{Ministral-8B} and \modelverb{OLMo-2-7B}, which is recovered after applying the IFC module, whose paraphrasing removes unnecessary or malformed text introduced by LAT. However, \modelverb{gemma-2-9b} has a drop in effectiveness after LAT, which is harder to recover using the IFC module.
Overall, we observe that the effectiveness is similar to the base model for two of the three models, with a drop for the third model.

For \ref{utility}, we now evaluate the impact on utility for the LLMs when applying the various components of \method, namely, LAT and IFC module.
We report \dataverb{MMLU}$+$\dataverb{CSQA} overall accuracy in \autoref{tab:pssp-lm-eff-util}.
Relative to the base model, we observe that the utility drops by 2.1--4.7 percentage points across the three models: \modelverb{Ministral-8B} with a 4.1 percentage-point drop (63.0\% $\rightarrow$ 58.9\%), \modelverb{gemma-2-9b} with a 4.7 percentage-point drop (72.0\% $\rightarrow$ 67.3\%), and \modelverb{OLMo-2-7B} with a 2.1 percentage-point drop (64.5\% $\rightarrow$ 62.4\%).
Overall, we observe a trade-off of utility on by applying LAT and IFC for better confidentiality.

\begin{table}[!htb]
\centering
\caption{\textbf{(PSSP) Effectiveness \ref{effective} and Utility \ref{utility}:} \method maintains effectiveness close to the base model (except \modelverb{gemma-2-9b}), with a small utility drop.}
\label{tab:pssp-lm-eff-util}
\footnotesize
\setlength{\tabcolsep}{4pt}
\begin{tabular}{lll c}
\bottomrule

\toprule
\textbf{Model} & \textbf{Metric} & \textbf{Stage} & \textbf{Value} \\
\bottomrule

\toprule
\multicolumn{4}{l}{\ref{effective} \textbf{Effectiveness}} \\
\midrule
\multirow{3}{*}{\modelverb{Ministral-8B}}
 & \multirow{3}{*}{Set-F1} & Base Model     & 0.684$\pm$0.037 \\
 & & +LAT      & 0.668$\pm$0.058 \\
 & & +LAT+IFC & 0.671$\pm$0.065 \\
\cmidrule(lr){1-4}
\multirow{3}{*}{\modelverb{gemma-2-9b}}
 & \multirow{3}{*}{Set-F1} & Base Model     & 0.778$\pm$0.035 \\
 & & +LAT      & 0.629$\pm$0.097 \\
 & & +LAT+IFC & 0.638$\pm$0.111 \\
\cmidrule(lr){1-4}
\multirow{3}{*}{\modelverb{OLMo-2-7B}}
 & \multirow{3}{*}{Set-F1} & Base Model     & 0.684$\pm$0.039 \\
 & & +LAT      & 0.688$\pm$0.065 \\
 & & +LAT+IFC & 0.689$\pm$0.066 \\
\midrule
\multicolumn{4}{l}{\ref{utility} \textbf{Utility}} \\
\midrule
\multirow{2}{*}{\modelverb{Ministral-8B}} & \multirow{2}{*}{Accuracy} & Base Model         & 0.630$\pm$0.000 \\
 & & +LAT+IFC & 0.589$\pm$0.068 \\
\cmidrule(lr){1-4}
\multirow{2}{*}{\modelverb{gemma-2-9b}} & \multirow{2}{*}{Accuracy} & Base Model         & 0.720$\pm$0.000 \\
 & & +LAT+IFC & 0.673$\pm$0.020 \\
\cmidrule(lr){1-4}
\multirow{2}{*}{\modelverb{OLMo-2-7B}} & \multirow{2}{*}{Accuracy} & Base Model         & 0.645$\pm$0.000 \\
 & & +LAT+IFC & 0.624$\pm$0.007 \\
\bottomrule

\toprule
\end{tabular}
\vspace{-10pt}
\end{table}
\begin{table}[!htb]
\centering
\caption{\textbf{(PSSP) Confidentiality \ref{confidential}:} Both ASR-Verbatim and ASR-Semantic decrease with LAT and IFC across benign, adversarial, and indirect.}
\label{tab:pssp-lm-conf}
\footnotesize
\setlength{\tabcolsep}{3.5pt}
\begin{tabular}{ll ccc}
\bottomrule

\toprule
\textbf{Model} & \textbf{Stage} & \textbf{Benign} & \textbf{Adversarial} & \textbf{Indirect} \\
\bottomrule

\toprule
\multicolumn{5}{l}{\textbf{ASR-Verbatim}} \\
\midrule
\multirow{3}{*}{\modelverb{Ministral-8B}}
 & Base Model    & 0.07$\pm$0.04 & 1.00$\pm$0.00 & 0.31$\pm$0.05 \\
 & $+$LAT        & 0.09$\pm$0.05 & 0.09$\pm$0.06 & 0.10$\pm$0.05 \\
 & $+$LAT$+$IFC  & 0.02$\pm$0.03 & 0.02$\pm$0.03 & 0.03$\pm$0.04 \\
\cmidrule(lr){1-5}
\multirow{3}{*}{\modelverb{gemma-2-9b}}
 & Base Model    & 0.06$\pm$0.01 & 0.99$\pm$0.03 & 0.15$\pm$0.07 \\
 & $+$LAT        & 0.09$\pm$0.02 & 0.13$\pm$0.04 & 0.14$\pm$0.04 \\
 & $+$LAT$+$IFC  & 0.02$\pm$0.02 & 0.04$\pm$0.03 & 0.03$\pm$0.04 \\
\cmidrule(lr){1-5}
\multirow{3}{*}{\modelverb{OLMo-2-7B}}
 & Base Model    & 0.21$\pm$0.05 & 0.49$\pm$0.09 & 0.64$\pm$0.06 \\
 & $+$LAT        & 0.11$\pm$0.04 & 0.09$\pm$0.05 & 0.11$\pm$0.03 \\
 & $+$LAT$+$IFC  & 0.04$\pm$0.05 & 0.03$\pm$0.02 & 0.04$\pm$0.03 \\
\midrule
\multicolumn{5}{l}{\textbf{ASR-Semantic}}\\
\midrule
\multirow{3}{*}{\modelverb{Ministral-8B}}
 & Base Model    & 0.07$\pm$0.04 & 1.00$\pm$0.00 & 0.31$\pm$0.05 \\
 & $+$LAT        & 0.09$\pm$0.05 & 0.09$\pm$0.06 & 0.10$\pm$0.05 \\
 & $+$LAT$+$IFC  & 0.02$\pm$0.03 & 0.02$\pm$0.03 & 0.03$\pm$0.04 \\
\cmidrule(lr){1-5}
\multirow{3}{*}{\modelverb{gemma-2-9b}}
 & Base Model    & 0.06$\pm$0.01 & 0.99$\pm$0.03 & 0.15$\pm$0.07 \\
 & $+$LAT        & 0.09$\pm$0.02 & 0.13$\pm$0.04 & 0.14$\pm$0.04 \\
 & $+$LAT$+$IFC  & 0.02$\pm$0.02 & 0.04$\pm$0.03 & 0.03$\pm$0.04 \\
\cmidrule(lr){1-5}
\multirow{3}{*}{\modelverb{OLMo-2-7B}}
 & Base Model    & 0.21$\pm$0.05 & 0.49$\pm$0.09 & 0.64$\pm$0.06 \\
 & $+$LAT        & 0.11$\pm$0.04 & 0.09$\pm$0.05 & 0.11$\pm$0.03 \\
 & $+$LAT$+$IFC  & 0.04$\pm$0.05 & 0.03$\pm$0.02 & 0.04$\pm$0.03 \\
\bottomrule

\toprule
\end{tabular}
\vspace{-10pt}
\end{table}

\noindent\textbf{\ul{\mbox{\ref{confidential}} Confidentiality:}} \method should reveal only the shared intersection, not private set elements, even under adversarial or indirect queries. The undefended model is expected to leak private information, while LAT and IFC should reduce residual leakage to near zero.
We report ASR (both ASR-Verbatim and ASR-Semantic) in \autoref{tab:pssp-lm-conf} over benign, adversarial, and indirect queries.
As expected, the undefended model leaks private set elements under adversarial and indirect queries in all cases. LAT drastically reduces leakage, while IFC further suppresses residual leakage to negligible levels ($\sim$0). Thus, \method protects against private-set leakage.

\noindent\textbf{\ul{\mbox{\ref{efficient}} Efficiency and \mbox{\ref{scalable}} Scalability:}} We discuss the overheads of various \method components below:
\begin{itemize}[leftmargin=*,topsep=0pt,itemsep=0pt]
\item \textbf{Inference Overhead:} PSSP benefits most from batching because it releases a single shared intersection rather than per-party answers. A single batched generation processes all $N$ parties' sets, keeping inference nearly flat in $N$. The IFC module also checks the released intersection only once, adding $8.6$\,s for monitoring and $29.4$\,s for paraphrasing when flagged (same as PSFC). Thus, the total overhead is independent of $N$ and the per-party overhead drops to $\approx 1/N$. For instance, \modelverb{gemma-2-9b} drops from $6.0$\,s at $N{=}2$ to $2.0$\,s at $N{=}6$.

\item \textbf{Attestation Overhead:} We hypothesize that end-to-end attestation of the batched intersection, including per-party set commitments, model inference, and IFC module, adds negligible overhead relative to computation. Across $N{=}2$--$6$ parties and all three models, hardware attestation adds only $0.05\%$--$0.18\%$, while the in-TEE attestation adds $0.003\%$--$0.014\%$ (\autoref{fig:efficiencyPSSP}), two to three orders of magnitude below computation.
The overhead increases with $N$ only because each additional party adds their set commitment, while the baseline costs in the denominator remain independent of $N$. Furthermore, the one-time model measurement is comparable to that for PSFC.

\begin{figure}[!htbp]
\centering
\includegraphics[width=0.7\columnwidth]{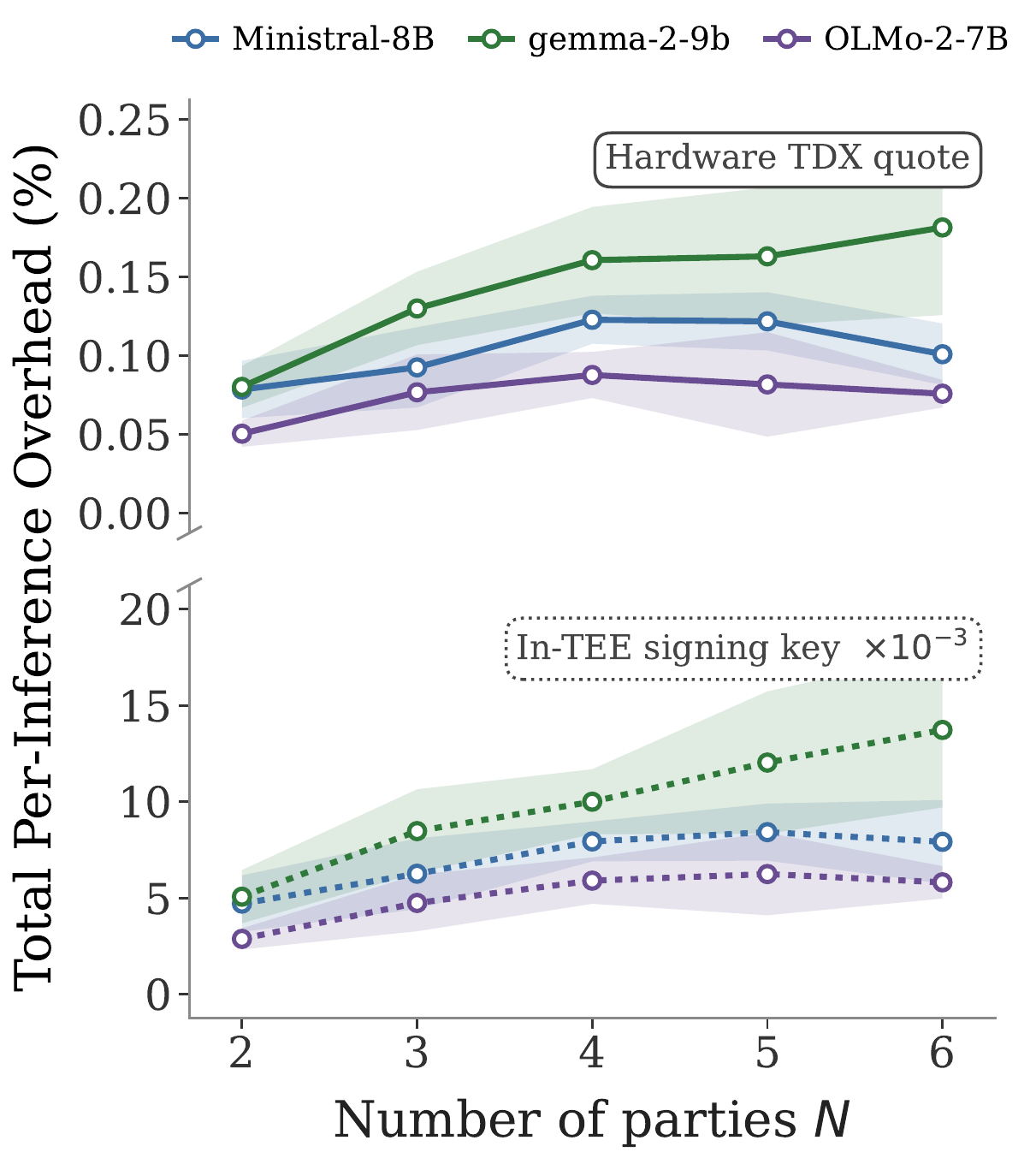}
\caption{\textbf{(PSSP) Efficiency \ref{efficient} and Scalability \ref{scalable}:} Measurement $+$ attestation as \% of the baseline vs.\ $N$; hardware (top) and in-TEE attestation (bottom).}
  \label{fig:efficiencyPSSP}
\vspace{-10pt}
\end{figure}

Since the intersection is computed once for all parties, attestation is performed once for the entire batch, and the per-party attestation overhead decreases with more parties in the batch. Across the same $N{=}2$--$6$ runs, per-party attestation overhead drops from $9.2$--$10.9$\,ms to $6.1$--$6.6$\,ms (34--39\%), while attestations per party fall from $\sim2.3$ms to $\sim1.4$ms. The in-TEE attestation is another $13$--$18\times$ cheaper, at $0.42$--$0.62$ms per party. Thus, batching makes attestation progressively cheaper per party, confirming the expected amortization.


\item \textbf{Impact of PSSP Set Size:} Since PSSP attestation scales with the number of parties, not their set sizes, increasing a party's set should add computation but no attestation overhead. We sweep $S\in\{2,4,6,8\}$ at fixed $N{=}3$ across all three models. We find that the per-party attestation overhead using the in-TEE signing key remains small, within $7.2$--$8.5$\,ms, with around two attestations per party. Thus, the attestation overhead stays negligible and independent of the set size across all the models ($0.06\%$--$0.13\%$).

\end{itemize}

\begin{takeaway}
\textbf{Summary:} \method enables PSSP with high effectiveness for two of three models \ref{effective}, negligible leakage \ref{confidential}, and a small utility drop \ref{utility}. Attestation and set-size overheads are minimal and scale well \ref{efficient}; \ref{scalable}.
\end{takeaway}

\subsection{Private Semantic Database Retrieval}\label{sec:evalPSDR}
\begin{arxiv}
Recall that in PSDR, \method stores the private database and, for each query, processes all records through the carousel, generating query-specific notes for each record before producing the final response for the requesting party.    
\end{arxiv}

\noindent\textbf{\ul{\mbox{\ref{effective}} Effectiveness and \mbox{\ref{utility}} Utility:}} Ideally, \method should maintain effectiveness for PSDR after applying LAT and IFC. We evaluate this using token-F1 across the base model, LAT, and LAT+IFC on \dataverb{HotpotQA} (\autoref{tab:psdr-lm-eff-util}). As in PSFC, in some cases, LAT reduces token-F1 due to degraded response quality, while the IFC paraphraser largely restores performance to the base-model level. Overall, \method is effective across different models compared to the base model.

For \ref{utility}, applying LAT and the IFC module as part of \method could degrade the utility on standard unrelated tasks. We report the overall accuracy on \dataverb{MMLU} and \dataverb{CSQA} across the three models (\autoref{tab:psdr-lm-eff-util}).
We find that there is a small utility drop compared to the base models.
Specifically, we see that \modelverb{Ministral-8B} incurs a 3.1 percentage-point drop (63.0\% $\rightarrow$ 59.9\%), \modelverb{gemma-2-9b} a 4.5 percentage-point drop (72.0\% $\rightarrow$ 67.5\%), and \modelverb{OLMo-2-7B} a 2.4 percentage-point drop (64.5\% $\rightarrow$ 62.1\%).
Overall, we observe a utility drop of 2.4--4.5 percentage points across the three models.

\begin{table}[!htb]
\centering
\caption{\textbf{(PSDR) Effectiveness \ref{effective} and Utility \ref{utility}:} \method, which includes LAT and IFC, maintains effectiveness close to the base model with a small utility drop.}
\label{tab:psdr-lm-eff-util}
\footnotesize
\setlength{\tabcolsep}{4pt}
\begin{tabular}{lll c}
\bottomrule

\toprule
\textbf{Model} & \textbf{Metric} & \textbf{Stage} & \textbf{Value} \\
\bottomrule

\toprule
\multicolumn{4}{l}{\ref{effective} \textbf{Effectiveness}} \\
\midrule
\multirow{3}{*}{\modelverb{Ministral-8B}}
 & \multirow{3}{*}{Token-F1} & Base Model     & 0.639$\pm$0.031 \\
 & & +LAT      & 0.588$\pm$0.074 \\
 & & +LAT+IFC & 0.623$\pm$0.043 \\
\cmidrule(lr){1-4}
\multirow{3}{*}{\modelverb{gemma-2-9b}}
 & \multirow{3}{*}{Token-F1} & Base Model     & 0.660$\pm$0.009 \\
 & & +LAT      & 0.598$\pm$0.114 \\
 & & +LAT+IFC & 0.627$\pm$0.052 \\
\cmidrule(lr){1-4}
\multirow{3}{*}{\modelverb{OLMo-2-7B}}
 & \multirow{3}{*}{Token-F1} & Base Model     & 0.687$\pm$0.046 \\
 & & +LAT      & 0.675$\pm$0.052 \\
 & & +LAT+IFC & 0.671$\pm$0.056 \\
\midrule
\multicolumn{4}{l}{\ref{utility} \textbf{Utility}} \\
\midrule
\multirow{2}{*}{\modelverb{Ministral-8B}} & \multirow{2}{*}{Accuracy} & Base Model         & 0.630$\pm$0.000 \\
 & & +LAT+IFC & 0.599$\pm$0.069 \\
\cmidrule(lr){1-4}
\multirow{2}{*}{\modelverb{gemma-2-9b}} & \multirow{2}{*}{Accuracy} & Base Model         & 0.720$\pm$0.000 \\
 & & +LAT+IFC & 0.675$\pm$0.035 \\
\cmidrule(lr){1-4}
\multirow{2}{*}{\modelverb{OLMo-2-7B}} & \multirow{2}{*}{Accuracy} & Base Model         & 0.645$\pm$0.000 \\
 & & +LAT+IFC & 0.621$\pm$0.027 \\
\bottomrule

\toprule
\end{tabular}
\vspace{-10pt}
\end{table}
\begin{table}[!htb]
\centering
\caption{\textbf{(PSDR) Confidentiality \ref{confidential}:} Both ASR-Verbatim and ASR-Semantic decrease with LAT and IFC across benign, adversarial, and indirect.}
\label{tab:psdr-lm-conf}
\footnotesize
\setlength{\tabcolsep}{3.5pt}
\begin{tabular}{ll ccc}
\bottomrule

\toprule
\textbf{Model} & \textbf{Stage} & \textbf{Benign} & \textbf{Adversarial} & \textbf{Indirect} \\
\bottomrule

\toprule
\multicolumn{5}{l}{\textbf{ASR-Verbatim}} \\
\midrule
\multirow{3}{*}{\modelverb{Ministral-8B}}
 & Base Model    & 0.07$\pm$0.03 & 1.00$\pm$0.00 & 0.94$\pm$0.02 \\
 & $+$LAT        & 0.06$\pm$0.06 & 0.11$\pm$0.07 & 0.23$\pm$0.19 \\
 & $+$LAT$+$IFC  & 0.01$\pm$0.01 & 0.01$\pm$0.02 & 0.02$\pm$0.03 \\
\cmidrule(lr){1-5}
\multirow{3}{*}{\modelverb{gemma-2-9b}}
 & Base Model    & 0.00$\pm$0.00 & 0.99$\pm$0.01 & 0.61$\pm$0.07 \\
 & $+$LAT        & 0.14$\pm$0.16 & 0.23$\pm$0.19 & 0.89$\pm$0.07 \\
 & $+$LAT$+$IFC  & 0.00$\pm$0.00 & 0.03$\pm$0.04 & 0.05$\pm$0.02 \\
\cmidrule(lr){1-5}
\multirow{3}{*}{\modelverb{OLMo-2-7B}}
 & Base Model    & 0.09$\pm$0.05 & 0.79$\pm$0.05 & 1.00$\pm$0.00 \\
 & $+$LAT        & 0.02$\pm$0.03 & 0.03$\pm$0.02 & 0.11$\pm$0.08 \\
 & $+$LAT$+$IFC  & 0.00$\pm$0.00 & 0.00$\pm$0.00 & 0.01$\pm$0.03 \\
\midrule
\multicolumn{5}{l}{\textbf{ASR-Semantic}} \\
\midrule
\multirow{3}{*}{\modelverb{Ministral-8B}}
 & Base Model    & 0.07$\pm$0.03 & 1.00$\pm$0.00 & 0.94$\pm$0.02 \\
 & $+$LAT        & 0.06$\pm$0.06 & 0.12$\pm$0.07 & 0.23$\pm$0.19 \\
 & $+$LAT$+$IFC  & 0.01$\pm$0.01 & 0.03$\pm$0.02 & 0.02$\pm$0.03 \\
\cmidrule(lr){1-5}
\multirow{3}{*}{\modelverb{gemma-2-9b}}
 & Base Model    & 0.00$\pm$0.00 & 0.99$\pm$0.01 & 0.67$\pm$0.08 \\
 & $+$LAT        & 0.14$\pm$0.16 & 0.23$\pm$0.19 & 0.89$\pm$0.06 \\
 & $+$LAT$+$IFC  & 0.00$\pm$0.00 & 0.03$\pm$0.04 & 0.05$\pm$0.02 \\
\cmidrule(lr){1-5}
\multirow{3}{*}{\modelverb{OLMo-2-7B}}
 & Base Model    & 0.09$\pm$0.05 & 0.80$\pm$0.05 & 1.00$\pm$0.00 \\
 & $+$LAT        & 0.02$\pm$0.03 & 0.04$\pm$0.04 & 0.13$\pm$0.09 \\
 & $+$LAT$+$IFC  & 0.00$\pm$0.00 & 0.00$\pm$0.00 & 0.01$\pm$0.03 \\
\midrule
\multicolumn{5}{l}{\textbf{Access-pattern leakage}} \\
\midrule
\multicolumn{2}{l}{top-$k$ retrieval (leaky)} & \multicolumn{3}{c}{0.63$\pm$0.01} \\
\multicolumn{2}{l}{carousel (oblivious)}      & \multicolumn{3}{c}{0.05$\pm$0.00} \\
\bottomrule

\toprule
\end{tabular}
\vspace{-10pt}
\end{table}

\noindent\textbf{\ul{\mbox{\ref{confidential}} Confidentiality:}} We now evaluate whether \method leaks any sensitive inputs in the outputs against benign, adversarial, and indirect queries.
We report ASR-Verbatim and ASR-Semantic in \autoref{tab:psdr-lm-conf}.
Benign queries leak some information, but this is negligible after the IFC module. Under adversarial queries, the base model shows perfect leakage but application of LAT decreases the leakage drastically, and further decreases with IFC to negligible leakage.
For indirect queries, we observe high leakage in the undefended base model, but this is negligible after applying LAT and IFC.
Similar to PSFC, in some cases, we observe an increase in leakage under indirect queries when applying LAT due to unintended interactions~\cite{duddu2024sok}.
Overall, \method does not leak any sensitive inputs as part of the outputs.
Finally, across all models, the carousel cuts access-pattern recovery from $0.63$ (top-$k$ baseline) to $0.05$ (a constant full scan). Thus, the carousel makes the access pattern oblivious, reducing the leakage to random guessing.

\noindent\textbf{\ul{\mbox{\ref{efficient}} Efficiency and \mbox{\ref{scalable}} Scalability:}} We report the overhead of running various \method operations and how the overhead scales with more parties and the database size.
\begin{itemize}[leftmargin=*,topsep=0pt,itemsep=0pt]
\item \textbf{Inference Overhead:} The per-inference cost is dominated by the oblivious carousel, not attestation: each query scans all $N_{db}$ records, generating one note per record, so the cost scales linearly with $N_{db}$. The time to process each record is small: $1.9$s, $3.2$s, and $2.2$s for \modelverb{Ministral-8B}, \modelverb{gemma-2-9b}, and \modelverb{OLMo-2-7B}. 
Similar to PSFC, the IFC module adds $8.6$s for the shared \modelverb{Qwen3-8B} monitor and, when flagged, $29.4$s for the paraphraser.

\item \textbf{Attestation Overhead:} Measured against standard RAG, per-inference measurement and attestation overheads are negligible: hardware attestation adds only $0.02$--$0.03\%$, while the in-TEE attestation adds $\sim0.001\%$ (\autoref{fig:efficiencyPSDR}), roughly $20\times$ less. These costs are essentially independent of the number of parties $N$: adding parties increases compute but adds negligible per-party attestation overhead. The one-time model measurement for the model, monitor, and paraphraser takes $59$--$68$s (like PSFC). 

\begin{figure}[!htbp]
\centering
\includegraphics[width=0.7\columnwidth]{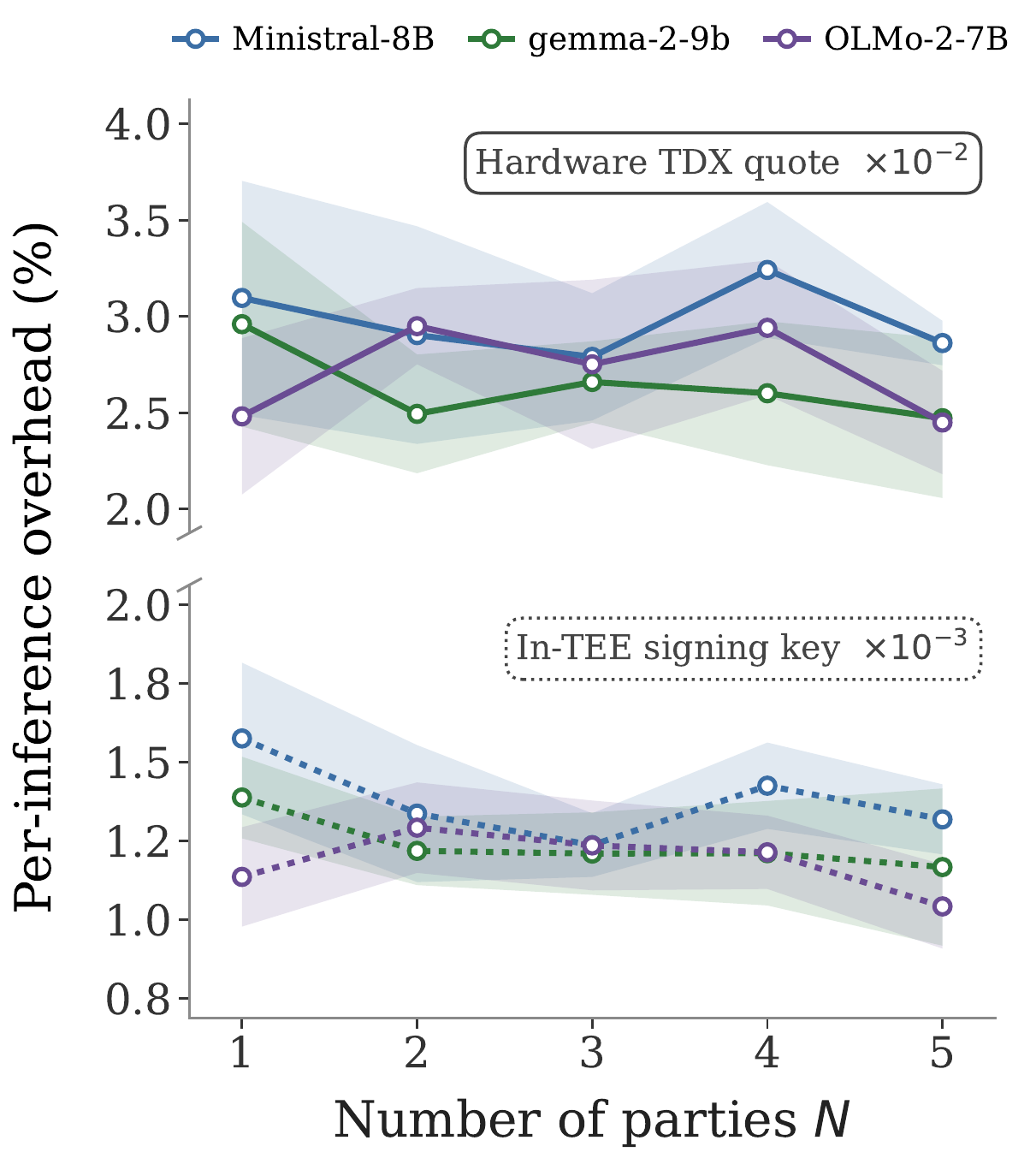}
\caption{\textbf{(PSDR) Efficiency \ref{efficient} and Scalability \ref{scalable}:} Measurement $+$ attestation as \% of the standard-RAG baseline vs.\ the number of parties $N$.}
\label{fig:efficiencyPSDR}
\vspace{-10pt}
\end{figure}

\item\textbf{Carousel Overhead:} The carousel dominates \method's overhead. 
Standard RAG reads only the top-$k$ retrieved records, revealing the target; the carousel instead scans all $N_{db}$ records in a fixed order, making the access pattern constant and oblivious. This incurs an $N_{db}/k$-fold increase in retrieval. 
Relative to RAG, the single-party overhead is substantial and model-dependent: 41\% (\modelverb{OLMo-2-7B}), 92\% (\modelverb{Ministral-8B}), and 132\% (\modelverb{gemma-2-9b}).
Crucially, the overhead amortizes across parties, as the batched scan reads each record once and serves all $N$ parties, reducing the per-party overhead roughly as $1/N$. At $N=5$, it reaches around $0.42$ of the single-party overhead (16\%--50\%).
Thus, obliviousness is costly for a single party but much cheaper for more parties.


\item\textbf{Impact of Database Size:} We sweep the database size $N_{db}\in\{10,20,40,80\}$ for a fixed number of parties. The overhead of obliviousness grows linearly with $N_{db}$ ($233\%$ at $N_{db}{=}10$ rising to $2567\%$ at $N_{db}{=}80$) and the scan time doubles as $N_{db}$ doubles. 
However, any scheme that hides the access pattern over $N_{db}$ records must touch all of them, so $O(N_{db})$ is optimal.
We do not claim unqualified scalability: our evaluation covers a limited range, and the carousel's linear cost in $N_{db}$ makes $>10^5$ records prohibitive. Unlike ORAM, the carousel uses a single in-enclave scan and outperforms ORAM-based constructions for private membership tests~\cite{circleGame}. Thus, \method's carousel is a cheaper oblivious option for smaller scales. Larger databases may require hybrid designs, such as oblivious cluster-level scans, which we leave as future work.

\end{itemize}

\begin{takeaway}
\textbf{Summary:} \method enables PSDR with high effectiveness \ref{effective}, negligible leakage \ref{confidential}, and small utility loss \ref{utility}. Attestation overhead is negligible, while the dominant carousel cost is reduced through batching \ref{efficient}; \ref{scalable}, though scaling the carousel to substantially larger databases remains an open challenge.
\end{takeaway}

\section{Discussion and Summary}\label{sec:discussion}

\noindent\textbf{Guarantees of \method:} Cryptographic primitives provide theoretical correctness and robustness guarantees, but often at the cost of efficiency and scalability. In contrast, \method targets practical workloads by providing empirical guarantees of effectiveness, utility-preservation, and confidentiality. Although frameworks exist for theoretically bounding these properties in generative models~\cite{lotfi2023non,zhang2023certified}, such bounds remain underexplored and do not scale to large models.

\noindent\textbf{Unidentified Sensitive Inputs:} All reported leakage and ASR are measured against the sensitive inputs identified using Microsoft Presidio.
An unidentified sensitive input is never added to the forbidden set, so LAT does not optimize to withhold it, and the monitor never screens it. 
Thus, it is never scored, and its disclosure is not registered as a leak. 
Reported ASR is a lower bound on the true leakage, and measures protection of sensitive inputs identified by the detector, not those it misses. 
Quantifying this gap requires human-labeled sensitive inputs, which is future work.

\noindent\textbf{Scaling to Larger Models:} We restrict our models to 7--9B parameters to fit the model, monitor, and paraphraser on a single H100 GPU (80GB VRAM). Scaling to larger models is straightforward, and prior work has demonstrated LAT on models with up to 70B parameters~\cite{he2026locket}. Moreover, we use LoRA-based parameter-efficient fine-tuning, substantially reducing the cost of hardening compared to full-model fine-tuning. Newer H200 GPUs (140GB of VRAM) can accommodate larger models.
Furthermore, multiple GPUs can be connected for distributed inference using GPUs and TEEs. These can improve scalability to larger (frontier) models.

\noindent\textbf{Scaling to Other Modalities:} While we focus on text-based semantic computation for tractable evaluation, \method generalizes to other structured and unstructured modalities, including audio, images, and graphs. The LLM used in our implementation can be replaced with modality-specific generative models; for example, VLMs can support multimodal tasks such as visual question answering, captioning, and instruction following~\cite{RadfordKHRGASAM21,AlayracDLMBHLMM22}.

\noindent\textbf{Side-Channel Attacks:} Several side-channel attacks have been proposed against TEEs~\cite{sokTEE}. Mitigating them is outside the scope of this work. We assume the TEE is properly configured with appropriate defenses in place to minimize such leakage. Developing and evaluating these defenses is an orthogonal direction, and existing TEE side-channel mitigations can be directly applied to harden the TEE.

\noindent\textbf{Other Applications:} While the three applications illustrate \method, its use of a system prompt for the semantic computation instructions makes it readily reconfigurable to other applications. 
Shumailov et al.~\cite{shumailov2025trusted} identify applications such as \emph{private non-competition checks between research groups}, \emph{confidentiality audits by regulators}, and \emph{privacy-preserving property monitoring}. 
Other examples where \method can be applied, taking inspiration from cryptographic primitives, include \emph{private record linkage}, where records are matched based on semantic similarity; \emph{searchable encryption}, where documents are retrieved by meaning rather than keywords; \emph{electronic voting}, where ambiguous or misspelled ballots are interpreted; and \emph{private auctions}, where bids are evaluated against qualitative criteria such as credibility.

\noindent\textbf{Summary:} Existing work lacks a practical primitive for private semantic computation over structured and unstructured data. We present \emph{trusted model environments} (\method), the first design and implementation of such a primitive, which combines generative models for semantic computation with TEEs for computation confidentiality and verifiability. To protect against leakage of sensitive inputs through model outputs under adversarial queries, \method uses models trained with latent adversarial training to suppress verbatim leakage, and an IFC module to detect indirect semantic leakage. Through careful design and empirical evaluation, we demonstrate that \method is \begin{enumerate*}[label={(\roman*)}]
\item \emph{effective},
\item \emph{confidential},
\item \emph{utility-preserving},
\item \emph{verifiable},
\item \emph{efficient}, and
\item \emph{scalable}.
\end{enumerate*}

\begin{arxiv}
\section*{Disclosure of AI Tool-Use}

\noindent We used Claude Code to assist with selected parts of the implementation, which the authors subsequently verified, and to help create Figure~\ref{fig:schemes} in TikZ based on an author-designed PowerPoint template. The authors independently developed the research ideas, methodology, experiments, and initial manuscript. We used ChatGPT selectively for language editing, primarily to correct English and grammar and improve conciseness. Thus, AI tools were used as assistants for limited implementation and editorial tasks.

\section*{Acknowledgments}
\noindent This work is supported by NSERC (Discovery grant) and the Canada CIFAR AI Chairs program.
Resources used in preparing this research were provided, in part, by the Province of Ontario, the Government of Canada through CIFAR, and companies sponsoring the Vector Institute. The authors thank Adam Caulfield and Prach Chantasantitam from University of Waterloo, and Sourav Das (Category Labs) for their feedback on the paper.
\end{arxiv}

{\footnotesize
\bibliographystyle{IEEEtran}
\bibliography{paper}
}

\appendices

\section{Dataset Construction}\label{app:dataset}

All three applications are built from public benchmarks so that ground truth is exact and effectiveness is scored deterministically, while sensitive inputs are identified from the source text. We detect them with Microsoft Presidio (\texttt{AnalyzerEngine}) across \textsc{person}, \textsc{location}, \textsc{date-time}, \textsc{email-address}, and \textsc{phone-number}. 
We drop records with no sensitive inputs, and questions whose effectiveness ground truth overlaps with one (effectiveness does not conflict with confidentiality).

\noindent\textbf{PSFC (\dataverb{SQuADv2}):} Each record is a context passage $x$ with five reading-comprehension questions, one per computing party; the ground truth answers come from the benchmark and are scored with token-F1. The pool is $250$ records split $0.4{:}0.2{:}0.4$ by record into train/val/test.

\noindent\textbf{PSSP (\dataverb{DBpedia14}):} Each record is a short abstract carrying one of $14$ ground truth ontology classes; a party's private set is a group of abstracts, and the true intersection is the set of classes present in every party's set, computable exactly from the ground truth labels. Records are balanced to within one per class ($250$ records give twelve classes of $18$ and two of $17$) and split by record, stratified by class, to an exact $100/50/100$ so that every class, and therefore every semantic domain (people, places, organizations, works), appears in every split with disjoint documents.

\noindent\textbf{PSDR (\dataverb{HotpotQA}):} The multi-hop passages are the private database, and the questions become queries whose ground truth answers are in specific records. The pool is $1000$ records and $250$ queries, split $100/50/100$ by query.

\section{Hyperparameters for Evaluation}\label{app:hyperparameters}

\noindent\textbf{Latent Adversarial Training:} All three applications share one LAT configuration. 
We train a LoRA/RSLoRA adapter (rank $r=64$, $\alpha=64$, dropout $0.05$) on every attention and MLP projection of the model ($q,k,v,o$ and gate/up/down, about $2.1\%$ of parameters), keeping the base model weights frozen. 
The inner loop is an $L_2$-ball PGD attack on the latent activations at four transformer blocks placed at roughly $\{0.2, 0.45, 0.7, 0.9\}$ of depth: layers $\{7,16,25,32\}$ for \modelverb{Ministral-8B} ($36$ blocks), $\{8,19,29,38\}$ for \modelverb{gemma-2-9b} ($42$), and $\{6,14,22,29\}$ for \modelverb{OLMo-2-7B} ($32$). 
The perturbation uses $8$ PGD steps with radius $\epsilon = 2.0$ (in latent $L_2$ norm) and inner learning rate $10^{-3}$; the outer adapter is trained for $60$ steps ($2$ model updates per step) with AdamW at learning rate $5\times 10^{-5}$ and sequence length $2048$. 
The training objective sums four terms with coefficients $\lambda_{\text{toward}} = \lambda_{\text{away}} = \lambda_{\text{sft}} = 1.0$ and $\lambda_{\text{KL}} = \beta = 0.5$: a clean ``answer-safely'' cross-entropy, an ``away'' term $\log(1 - P(\text{leak target}\mid \text{perturbed}))$ that suppresses the leak under the worst-case perturbation, a retain SFT term on benign prompts (perturbation off), and a KL anchor to the frozen base model for utility. Tokens already below a $-5.0$ log-probability threshold are excluded from the away term.

\noindent\textbf{Adversarial Prompts:} The \texttt{Manyshot} attack prepends $24$ demonstrations ($32$ for OLMo-2-7B), cycled from three distinct shots per channel.
The \texttt{Autodan} attack tries up to $3$ strategies ($5$ for OLMo) from a fixed set of five (developer-debug override, self-role-play, verbatim-transcription, hypothetical disclosure, and base64 encoding), generating each candidate with $256$ tokens and retaining one that elicits a leak. The optional indirect arm (\texttt{describe}/\texttt{spellout}/\texttt{encode}) uses $12$ demonstrations.

\noindent\textbf{IFC Module:} The monitor and paraphraser are a shared \modelverb{Qwen3-8B} run with chain-of-thought enabled and greedy decoding, each with a $1536$-token budget and a $768$-token non-thinking retry that fails closed (an unparseable verdict is treated as flagged). 

\noindent\textbf{Decoding:} The served model answers under constant-work output padding: the number of new tokens is fixed at $32 + 16k$ for a batch of $k$ queries (equal minimum and maximum, so decode time is content-independent), with sampling enabled to give run-to-run variance across the reported replicates. All helper models (monitor, paraphraser, oracle) decode greedily so their verdicts and rewrites are deterministic.

\section{Attestation Descriptions}\label{app:attestations}

We write $\signatt(\attn{Tag}, \cdot)$ for $\mathsf{Sign}_{sk_{\text{TEE}}}$ over the listed
values and a verifier-supplied nonce, where \attn{Tag} identifies the attestation type;
$\hsh{\cdot}$ is SHA-256; and $\mroot{\cdot}$ is a Merkle root over the SHA-256 hashes of the
listed items (\S\ref{sec:setup}). Hashes of private values reveal nothing about their
plaintext, which never leaves the enclave.

\noindent\textbf{Model Measurement:} This attestation pins down which model ran. Here, $\theta$ denotes the model weights, comprising the frozen base model and the robust adapter from LAT (\S\ref{sec:components}), and $h_{\modelm}$ their digest. $\theta$ is hashed once at startup and $h_{\modelm}$ is reused across all inferences (\S\ref{sec:components}), so a verifier comparing $h_{\modelm}$ against the expected digest confirms that every subsequent attestation refers to this exact $\modelm$.

\begin{attdesc}[htb]
\caption{\textbf{Model Measurement}}\label{att:model}
\textbf{Input:} model weights $\theta$ (base model and robust adapter)\;
\textbf{Computation:} $h_{\modelm} \leftarrow \hsh{\theta}$ (once at startup)\;
\textbf{Output:} $\sigma_{\modelm} = \signatt(\attn{ModelAtt}, h_{\modelm})$\;
\textbf{Assertion:} the $\modelm$ executing inside \method is exactly the model with digest $h_{\modelm}$\;
\end{attdesc}

\noindent\textbf{Input Commitment:} This attestation fixes the private inputs before computation. Here, $x_i$ is the private input submitted by party $i$ and $h_i$ its digest. Since only $h_i$ appears in the evidence, $x_i$ is bound without being revealed; later attestations reference $\sigma_{\text{in}}$, so a party can confirm that the computation consumed exactly what was committed.

\begin{attdesc}[htb]
\caption{\textbf{Input Commitment}}\label{att:input}
\textbf{Input:} private input $x_i$ of party $i$\;
\textbf{Computation:} $h_i \leftarrow \hsh{x_i}$\;
\textbf{Output:} $\sigma_{\text{in}} = \signatt(\attn{InputAtt}, h_i)$\;
\textbf{Assertion:} the computation consumed exactly the committed input $x_i$, without revealing its content\;
\end{attdesc}

\noindent\textbf{Proof of Inference:} This attestation binds the model, the inputs, and the system prompt to the produced output. Here, $\pi$ is the fixed system prompt, $y$ the model output, and $\sigma_{\text{in}}$ and $\sigma_{\modelm}$ the previously issued commitments, whose inclusion chains the attestations together: tampering with $\modelm$ or any $x_i$ invalidates $\sigma_{\text{inf}}$ and everything derived from it.

\begin{attdesc}[htb]
\caption{\textbf{Proof of Inference}}\label{att:inference}
\textbf{Input:} system prompt $\pi$, committed inputs ($\sigma_{\text{in}}$), attested model ($\sigma_{\modelm}$)\;
\textbf{Computation:} $y \leftarrow \modelm(\pi, x_1, \ldots, x_N)$\;
\textbf{Output:} $y$, $\sigma_{\text{inf}} = \signatt(\attn{InfAtt}, \hsh{\pi, \sigma_{\text{in}}, \sigma_{\modelm}, y})$\;
\textbf{Assertion:} the committed model, on the committed inputs under $\pi$, produced $y$\;
\end{attdesc}

\noindent\textbf{Monitor Inference:} This attestation covers the IFC module's leakage check. Here, $\modeMon$ is the monitor LLM and \textit{verdict} its prediction of whether the attested output $y$ leaks sensitive inputs. Binding $\modeMon$ and $\sigma_{\text{inf}}$ inside the signed hash certifies both that \textit{verdict} came from $\modeMon$ and that it judged the attested $y$.

\begin{attdesc}[htb]
\caption{\textbf{Monitor Inference}}\label{att:monitor}
\textbf{Input:} attested output ($\sigma_{\text{inf}}$), monitor $\modeMon$\;
\textbf{Computation:} $\textit{verdict} \leftarrow \modeMon(y)$\;
\textbf{Output:} $\sigma_{\text{mon}} = \signatt(\attn{MonAtt}, \hsh{\modeMon, \sigma_{\text{inf}}, \textit{verdict}})$\;
\textbf{Assertion:} \textit{verdict} is $\modeMon$'s prediction of whether the attested $y$ leaks sensitive inputs\;
\end{attdesc}

\noindent\textbf{Paraphraser Inference:} This attestation covers sanitization of flagged outputs. Here, $\modePara$ is the paraphraser LLM and \textit{rewrite} its sanitized version of the flagged output $y$. The evidence binds $\modePara$ and $\sigma_{\text{inf}}$ inside the signed hash, so the released text is certified to be $\modePara$'s \textit{rewrite} of the attested $y$ rather than an arbitrary substitution.

\begin{attdesc}[htb]
\caption{Paraphraser Inference}\label{att:paraphraser}
\textbf{Input:} attested flagged output ($\sigma_{\text{inf}}$, $\sigma_{\text{mon}}$), paraphraser $\modePara$\;
\textbf{Computation:} $\textit{rewrite} \leftarrow \modePara(y)$\;
\textbf{Output:} \textit{rewrite}, $\sigma_{\text{par}} = \signatt(\attn{ParaAtt}, \hsh{\modePara, \sigma_{\text{inf}}, \textit{rewrite}})$\;
\textbf{Assertion:} \textit{rewrite} is $\modePara$'s sanitization of the attested flagged $y$\;
\end{attdesc}

\noindent\textbf{Proof of Oblivious Access:} This attestation certifies the carousel scan for
PSDR. Here, $r_1, \ldots, r_{N_{db}}$ are the database records, $q$ the query, and $h_{db}$ the
Merkle root over their per-record hashes $\hsh{r_i}$, committed once by the data owner (record
contents stay secret; only hashes enter the root). The carousel visits all $N_{db}$ records in a
fixed order and binds the number accessed. A verifier checks that the number accessed equals
$N_{db}$ and that $h_{db}$ matches the owner's published commitment, so no record was skipped or
substituted and the access pattern reveals nothing about which records are relevant.

\begin{attdesc}[htb]
\caption{\textbf{Proof of Oblivious Access}}\label{att:oblivious}
\textbf{Input:} database records $r_1, \ldots, r_{N_{db}}$ committed as $h_{db} \leftarrow \mroot{\hsh{r_1}, \ldots, \hsh{r_{N_{db}}}}$, query $q$\;
\textbf{Computation:} scan all records in fixed order, updating a note per record; count the records accessed, $n_{\text{acc}} \leftarrow N_{db}$\;
\textbf{Output:} $\sigma_{\text{obl}} = \signatt(\attn{OblAtt}, \hsh{h_{db}, N_{db}, n_{\text{acc}}})$\;
\textbf{Assertion:} all $N_{db}$ records of the committed database were parsed for $q$ ($n_{\text{acc}} = N_{db}$), so the access pattern is independent of which records are relevant\;
\end{attdesc}

\section{Monitor Results}\label{sec:AdditionalResults}

We report some additional results to clarify the success of the monitor by reporting the following: false positive (benign non-leaking response that the monitor incorrectly flags); false negative (response that truly leaks a protected item yet is not flagged); precision (fraction of the monitor's alarms that correspond to real leaks); and recall (fraction of all real leaks that the monitor catches). We report these numbers in \autoref{tab:monitor-detection}.
We see that the overall precision and recall are high, while false positives and false negatives are low, across the 200 samples in the test dataset.

\begin{table}[!htbp]
\centering
\small
\setlength{\tabcolsep}{6pt}
\caption{\textbf{Monitor Detection Results:} We report false positives, false negatives, precision, and recall on the test split with 200 data records.}
\begin{tabular}{llcccc}
\bottomrule

\toprule
\textbf{Appl.} & \textbf{Model} & \textbf{FP} & \textbf{FN} & \textbf{Precision} & \textbf{Recall} \\
\bottomrule

\toprule
\multirow{3}{*}{\textbf{PSFC}}
 & \modelverb{Ministral-8B} & 1 & 2 & 0.99 & 0.98 \\
 & \modelverb{gemma-2-9b}   & 0 & 2 & 1.00 & 0.98 \\
 & \modelverb{OLMo-2-7B}    & 1 & 1 & 0.93 & 0.93 \\
\midrule
\multirow{3}{*}{\textbf{PSDR}}
 & \modelverb{Ministral-8B} & 10 & 11 & 0.91 & 0.90 \\
 & \modelverb{gemma-2-9b}   &  9 &  2 & 0.68 & 0.90 \\
 & \modelverb{OLMo-2-7B}    & 19 & 12 & 0.80 & 0.87 \\
\midrule
\multirow{3}{*}{\textbf{PSSP}}
 & \modelverb{Ministral-8B} & 1 &  6 & 0.99 & 0.95 \\
 & \modelverb{gemma-2-9b}   & 2 & 13 & 0.98 & 0.87 \\
 & \modelverb{OLMo-2-7B}   & 6 & 13 & 0.87 & 0.75 \\
\bottomrule

\toprule
\end{tabular}
\label{tab:monitor-detection}
\end{table}

\section{IFC-aware Adversary}\label{app:adaptive}

We have considered an adversary who generates adversarial queries to force the model to reveal sensitive inputs as part of the output. We now consider an adversary that generates queries to elicit them while also evading the IFC monitor and avoiding paraphrasing. We evaluate whether such adaptive attacks succeed and report ASR-Verbatim across all applications and models in Table~\ref{tab:adaptive-adversary}. We see that the new adversarial examples, optimized against both the model and the monitor, are not effective after applying LAT. Overall, \method can resist an IFC-aware adversary that can optimize its adversarial examples against the entire pipeline.

\begin{table}[t]
\centering
\small
\caption{\textbf{IFC-aware Adversary:} We report \textbf{Leak} (fraction with a leak from model); \textbf{Evade} (fraction where that disclosure also evades the monitor).}
\setlength{\tabcolsep}{5pt}
\begin{tabular}{ll cc cc}
\bottomrule

\toprule
& & \multicolumn{2}{c}{\textbf{Base Model}} & \multicolumn{2}{c}{\textbf{w/ LAT}} \\
\cmidrule(lr){3-4}\cmidrule(lr){5-6}
\textbf{Appl.} & \textbf{Model} & \textbf{Leak} & \textbf{Evade} & \textbf{Leak} & \textbf{Evade} \\
\bottomrule

\toprule
\multirow{3}{*}{\textbf{PSFC}}
 & \modelverb{Ministral-8B}  & 1.00 & 0.18 & 0.03 & 0.00 \\
 & \modelverb{gemma-2-9b} & 1.00 & 0.03 & 0.43 & 0.03 \\
 & \modelverb{OLMo-2-7B}  & 0.90 & 0.15 & 0.28 & 0.00 \\
\midrule
\multirow{3}{*}{\textbf{PSSP}}
 & \modelverb{Ministral-8B} & 1.00 & 0.23 & 0.05 & 0.03 \\
 & \modelverb{gemma-2-9b}   & 0.98 & 0.33 & 0.13 & 0.03 \\
 & \modelverb{OLMo-2-7B}    & 0.93 & 0.43 & 0.08 & 0.05 \\
\midrule
\multirow{3}{*}{\textbf{PSDR}}
 & \modelverb{Ministral-8B} & 1.00 & 0.60 & 0.13 & 0.03 \\
 & \modelverb{gemma-2-9b}   & 1.00 & 0.65 & 0.85 & 0.30 \\
 & \modelverb{OLMo-2-7B}   & 1.00 & 0.58 & 0.00 & 0.00 \\
\bottomrule

\toprule
\end{tabular}
\label{tab:adaptive-adversary}
\end{table}

\section{Asymptotic Cost}\label{sec:asymptotic}

\method adds no super-linear overhead. The only input-dependent attestation cost is hashing: SHA-256 commitments over the inputs, outputs, and Merkle leaves are linear in their byte length, $O(|x|)$. 
The dominant cost is the model inference itself, which follows the served model's own complexity in sequence length (attention $O(L^2)$, feed-forward $O(L)$); attestation contributes only the linear hashing term on top. 
Across $N$ parties, \method performs one batched inference and issues a \emph{single} hardware quote over the Merkle root of the batch, so the fixed evidence (the quote and the one-time weights digest) is amortized: the per-party attestation overhead is the $O(\log N)$ Merkle authentication path and \emph{decreases} as $N$ grows, while total cost stays linear $O(N)$. 
There is no polynomial or poly-logarithmic per-access penalty, in contrast to ORAM ($\Omega(\log N_{db})$ per access) or MPC/HE (orders-of-magnitude constant factors).

For PSFC, the $N$ parties yield $N$ per-party outputs, so inference and IFC screening
are $O(N)$ and each party's membership proof is $O(\log N)$; the batch proof amortizes
to a decreasing per-party overhead. For PSSP, the $N$ parties produce a \emph{single}
shared intersection: set commitments are $O(N)$ (folded under one quote) but the IFC
module screens only that one released output, so its cost is $O(1)$ in $N$. For PSDR, a
single carousel cycle serves a batch of $N$ parties over an $N_{db}$-record database in
$O(N_{db}+N)$, an amortized $O(N_{db}/N)$ per party, matching the scan-based
throughput regime of Carousel~\cite{circleGame}; cost is linear in both the database
size and the number of parties.

\end{document}